\documentclass[10pt,journal,letterpaper,onecolumn]{IEEEtran} 
\ifCLASSOPTIONcompsoc
\else
\fi
\ifCLASSINFOpdf
\else
\fi

\usepackage{atbegshi}
\AtBeginDocument{\AtBeginShipoutNext{\AtBeginShipoutDiscard}}
\usepackage{amssymb,amsmath}
\usepackage{graphicx}
\usepackage{subfigure}
\usepackage{multicol}
\usepackage{multirow}
\usepackage{rotating}
\usepackage{algorithmic}
\usepackage{algorithm}
 \usepackage{amsthm}

\begin{document}
%

\title{An Optimized Information-Preserving Relational Database Watermarking Scheme for Ownership Protection of Medical Data}\thanks{A shortened version of this work was published in IEEE Transactions on Knowledge and Data Engineering in 2012.
}

\author{M.~Kamran and \IEEEmembership{}Muddassar~Farooq~\IEEEmembership{}
\IEEEcompsocitemizethanks{\IEEEcompsocthanksitem M. Kamran is with Center for Research in Data Mining and Data Security (CRDMDS), Department of Computer Science, COMSATS Institute of Information Technology Wah Campus, Wah Cantt,
Pakistan.\protect\\
E-mail: muhammad.kamran@ciitwah.edu.pk\IEEEcompsocthanksitem Muddassar
Farooq is with Next Generation Intelligent
Networks Research Center (nexGIN RC) Pakistan.
\protect\\E-mail: muddassar.farooq@gmail.com}
\thanks{}}
\markboth{}
{Shell \MakeLowercase{\textit{et al.}}: Bare Demo of IEEEtran.cls
for Computer Society Journals}

\IEEEcompsoctitleabstractindextext{%
\begin{abstract}
Recently, a significant amount of interest has been developed in
motivating physicians to use e-health technology (especially
Electronic Medical Records (EMR) systems). An important utility of
such EMR systems is: a next generation of Clinical Decision Support
Systems (CDSS) will extract knowledge from these electronic medical
records to enable physicians to do accurate and effective diagnosis.
It is anticipated that in future such medical records will be shared
through cloud among different physicians to improve the quality of
health care. Therefore, right protection of medical records is
important to protect their ownership once they are shared with third
parties. Watermarking is a proven well known technique to achieve
this objective. The challenges associated with watermarking of EMR
systems are: (1) some fields in EMR are more relevant in the
diagnosis process; as a result, small variations in them could
change the diagnosis, and (2) a misdiagnosis might not only result
in a life threatening scenario but also might lead to significant
costs of the treatment for the patients. The major contribution of
this paper is an information-preserving watermarking scheme to
address the above-mentioned challenges. We model the watermarking
process as a constrained optimization problem. We demonstrate,
through experiments, that our scheme not only preserves the
diagnosis accuracy but is also resilient to well known attacks for
corrupting the watermark. Last but not least, we also compare our
scheme with a well known threshold-based scheme to evaluate relative
merits of a classifier. Our pilot studies reveal that -- using
proposed information-preserving scheme -- the overall classification
accuracy is never degraded by more than 1\%. In comparison, the
diagnosis accuracy, using the threshold-based technique, is degraded
by more than 18\% in a worst case scenario.
\end{abstract}

}

\maketitle

\IEEEdisplaynotcompsoctitleabstractindextext

\IEEEpeerreviewmaketitle

\section{Introduction}

\IEEEPARstart{P}{ervasive} and ubiquitous deployment of information
and communication technology infrastructure is bringing a revolution
the way health care services are provided. Consequently, e-health
technology and its associated systems are being actively
standardized, adopted and deployed. An important component of modern
e-health technology is EMR systems in which a patient's medical
history, his vitals, lab tests, and diagnostic images are stored
\cite{feldstein2006electronic}. In the health reform package, Obama
administration is offering between \$44,000 and \$64,000 to the
physicians who would use EMR systems. The incentive amount for
adopting EMR systems at a hospital is \$11 million
\cite{ObamaIncentive}.

In order to provide quality care, it is relevant that the access to
medical records be provided in a ubiquitous fashion to the concerned
physicians. Therefore, the inter-operable EMR systems are becoming
the top priority in the US \cite{InterOperablityUSPriority}. This
demands exchanging and sharing sensitive health records in a network
cloud. Moreover, next generation of CDSSs will have the ability to
extract knowledge from the electronic medical records and use them
to assist physicians in making accurate and effective diagnosis. In
such scenarios, the shared data might be illegally sold to third
parties by an unauthorized party. In order to cater for such a
situation, the data needs to be right protected so that an
unauthorized party might be sued in a court of law. This is only
possible, if the data owner is able to prove that the illegally sold
data is his property. Therefore, it is important to not only protect
the privacy of the patients \cite{alhaqbani2009privacy},
\cite{peleg2008situation}, \cite{benaloh2009patient} but also the
ownership (copyright) of the medical data shared with collaborative
partners or third party vendors. In a recent survey, it is reported
that frauds related to stolen medical records have risen from 3$\%$
in 2008 to $7\%$ in 2009 (approximately a $112$\% increase)
\cite{fraudIncrease}. Similarly, confidential medical records of
patients in the EMR of a prestigious private hospital in UK were
illegally sold to undercover investigators for $4\pounds$ per record
\cite{IllegalEMRSale}. It is stated that theft of medical records is
more serious because it takes more than twice the time to detect a
fraud related to the medical information and the average cost is
\$12,100 which is more than twice the cost for other types of data
theft \cite{EMRDataCost}. A recent case related to an illegal sale
of the medical data is reported in \cite{patientssue}. Therefore, it
is important that medical records be right protected in a manner
where ownership could unambiguously be determined. (The data theft
is becoming a serious issue even for the most reliable brands like
LexisNexis, Polo Ralph Lauren, HSBC, NCR, and a number of renowned
universities \cite{famousCampaniesDataTheft}).

In this paper, we propose a novel right protection scheme that will
establish the ownership of EMR data, and consequently, will make its
illegal sale very difficult (even if the intruder has altered the
original data). Recently, some watermarking techniques are proposed
for databases \cite{kiernan2002watermarking}, \cite{sion2004rights},
\cite{shehab2008watermarking} but they are unable to address a
significant challenge related to EMR: \textit{insertion of a
watermark must not result in changing health and medical history of
a patient to a level where a decision maker (or system) can
misdiagnose the patient.} If a patient is misdiagnosed, it might not
only put his life on risk but also result in significantly enhancing
the cost of health care. In order to address this problem, we
propose the concept of \textit{information-preserving watermarking}.
The basic motivation behind this technique is to first develop a
model that determines the correlation of different features with the
diagnosis. We model the process of computing a suitable watermark as
a constrained optimization problem \cite{agrawal2003watermarking}.
Once the watermark is computed, it is inserted (with $O(n)$
complexity) by utilizing the knowledge of correlation of those
features that have negligible impact on the diagnosis. As a result,
the diagnosis rules are preserved and it is important because these
rules are used for diagnosing the patients and suggesting a relevant
treatment plan. Consequently, if a rule is changed due to insertion
of watermark then the suggested treatment would also change which
may cause serious harm to the life of a patient. Moreover, wrong
treatment will also waste time and money of the patients. Therefore,
it is mandatory that during watermarking of EMR, diagnosis rules
must be preserved. Moreover, the inserted watermark should be
imperceptible to intruders and they should not be able to corrupt it
by launching malicious attacks \cite{agrawal2003watermarking},
\cite{pournaghshband2008new}. Last but not least, multiple
insertions of the same watermark should not corrupt medical records.

Recent research proves that computational intelligence techniques,
especially Particle Swarm Optimization (PSO)
\cite{hassan2005comparison}, \cite{zitzler2000comparison}, are
ideally suited for solving constrained optimization problems in
realtime. In this paper, we use PSO to create an optimized watermark
that -- once inserted into an EMR -- does not alter the diagnosis
rules. The major contributions of our work are: (1) an intelligent
information-preserving watermarking technique for EMR systems that
ensures data usability constraints and also preserves the rules
after the watermark encoding, (2) realtime insertion of watermark
into an EMR system, and (3) a robust watermark decoding scheme that
is resilient against all kinds of malicious attacks.

The rest of the paper is organized as follows. In the next section,
we provide a brief overview of the research related to our work by
emphasizing the different direction of our work. In section
\ref{sec:ProposedMethod}, we discuss our proposed scheme followed by
experiments and results in section \ref{sec:ExperimentsandResults}.
Finally, we conclude the paper with an outlook to our future
research.

\section{Related Work}
\label{sec:RelatedWork}

Agrawal et al. \cite{agrawal2003watermarking} proposed the idea of
watermarking using least significant bits (LSB). They do not account
for multibit watermarks which makes their technique vulnerable
against simple attacks, for example shifting of only one least
significant bit results in loss of watermark. Xinchun et al. have
proposed a watermarking scheme for relational databases that uses
weights (assigned by users) of attributes to identify the location
where the watermark is to be inserted \cite{ cui2007weighted}. They
intuitively argue that the primary key of a database is the most
important attribute and hence a watermark should not be inserted in
it. They, however, did not consider the importance of ranking
attributes because their focus was not on using a database for data
mining and decision support systems.

Sion et al. \cite{sion2004rights} presented a watermarking scheme in
which data is partitioned using marker tuples. But if an attacker
launches an attack on marker tuples, the synchronization is lost.
The marker tuples decide the boundaries for the partitions but
successful insertion or deletion attacks on marker tuples would
result in synchronization errors.  Li et al. \cite{li2004tamper}
have proposed a watermarking technique to detect and localize
alterations made to a database relation with relevant attributes. Li
et al. have proposed a technique for fingerprinting relational data
\cite{li2005fingerprinting} by extending the technique proposed in
\cite{agrawal2003watermarking}. Jiang et al. proposed an invisible
watermarking technique using Discrete Wavelet Transform (DWT) in
\cite{jiang2009watermarking}.

A well known problem with all these techniques is: they are not
resilient to malicious attacks launched by an intruder. Recently,
Shehab et al. have proposed a robust watermarking scheme in
\cite{shehab2008watermarking}. They model a bit encoding algorithm
as an optimization problem and use Genetic Algorithm (GA)
\cite{holland1992genetic} and Pattern Search (PS)
\cite{lewis1998pattern} to do the optimization. They use a threshold
based approach coupled with a majority vote to decode the watermark.
The focus of their work is to show that their scheme is robust to
malicious attacks (insertion and deletion) launched by an intruder.
In \cite{bertino2005privacy} Bertino et al. have proposed a
technique for preserving the ownership of outsourced medical data
but they their focus was on preserving the classification potential
and the diagnostic rules.

In comparison, the focus of our work is to rank attributes
(features) on the basis of their importance in a decision making
process. Our objective is to identify weak attributes by developing
a knowledge model that correlates the effect of an attribute on the
decision making process. The major difference of our approach from
previous techniques is that we first identify the classification
potential of each attribute for a given diagnosis. Furthermore, we
use the knowledge of classification potential for every candidate
attribute to calculate the watermark which ensures that the data
usability constraints remain intact and also the diagnosis rules are
preserved. As a result, we meet the most important requirement while
right protecting an EMR: to preserve the diagnostic semantics of the
health record of a patient (i.e. a patient should not be
misdiagnosed).

The proposed technique uses an optimization technique to first
create a watermark which ensures that data usability constraints are
not violated. Once the watermark is created, we embed it in realtime
into an EMR system. In comparison, other techniques
\cite{shehab2008watermarking} use optimization techniques during
watermark embedding and hence they take a significant amount of
time. As a consequence, they are not suitable for realtime insertion
of watermark into an EMR system.

Moreover, our technique is also robust to malicious attacks because
it does not target any specific group of tuples \footnote{Throughout
this text -- unless otherwise specified -- we use the terms tuples,
rows and records to refer to rows in $EMR$.} for watermarking
(almost all other techniques developed so far use a \emph{secret
key} to select some group of tuples as candidate attribute(s) for
watermarking).

\section{The Proposed Information-Preserving Watermarking Technique}
\label{sec:ProposedMethod}

\begin{figure*}[htp]\scriptsize
\centering \includegraphics[angle=0,
width=1\columnwidth]{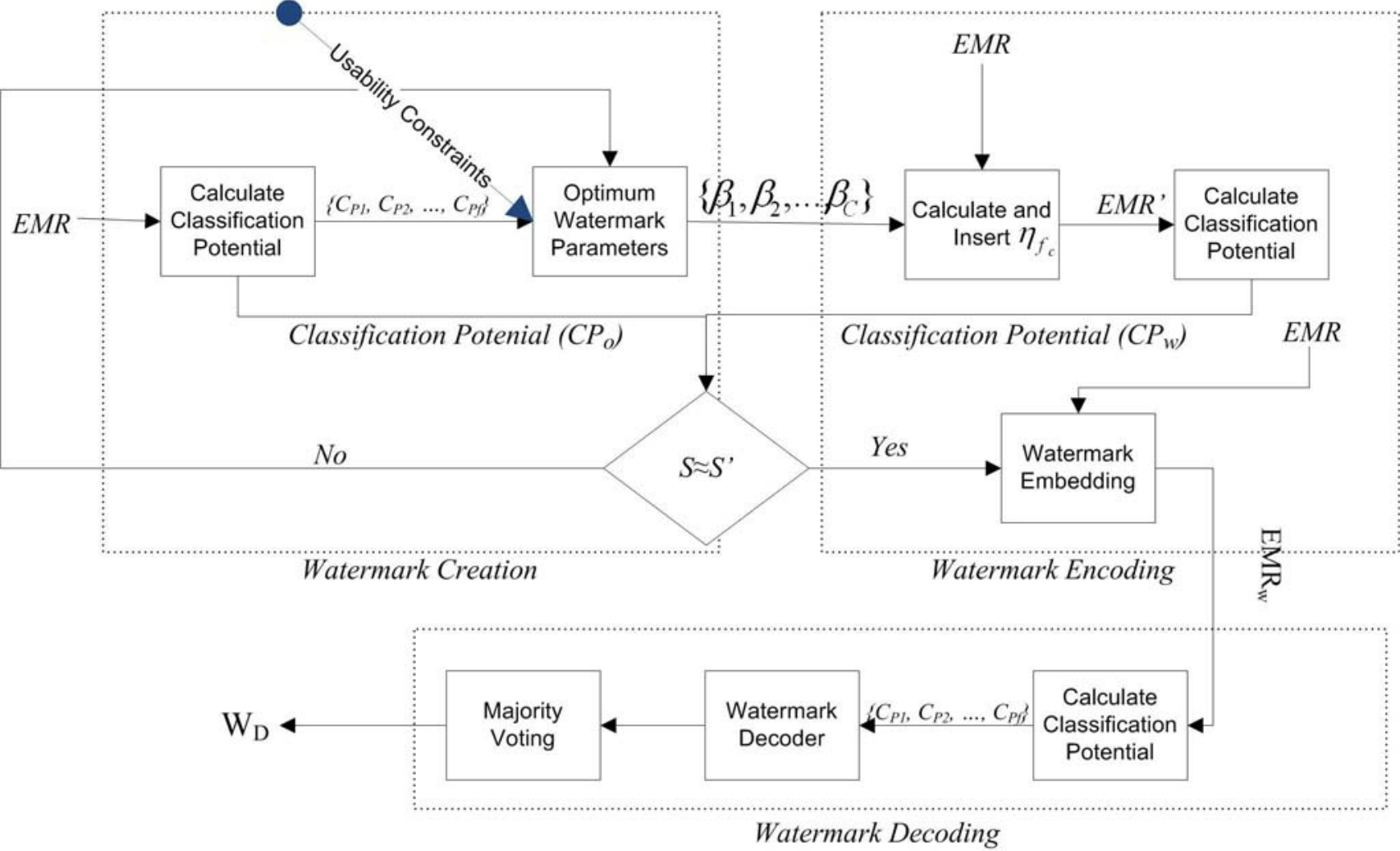}
\caption{The architecture of the proposed
scheme.}\label{fig:PaperArchitecture}
\end{figure*}

We propose a framework that does information-preserving right
protection of EMR systems (see Figure \ref{fig:PaperArchitecture}).
The framework operates in two modes: (1) Watermark encoding, and (2)
Watermark decoding. In the watermark encoding phase, the objective
is to determine a watermark that, once inserted into an
$EMR$\footnote{Throughout this text -- unless otherwise specified --
we use the terms database, dataset and $EMR$ to refer to the
$EMR$.}, does not alter the vital features to an extent that the
patient is misdiagnosed. Similarly, the objective of the decoding
phase is to accurately detect a watermark in an efficient manner. We
now discuss the watermark encoding scheme in the following.

\subsection{Watermark Encoding Phase}
 \label{sec:DeterminationofWatermarkParameters}

In this phase (or mode), three important steps are to be performed:
(1) quantifying the importance of a patient's vitals (or history
information) and their correlation with the diagnosis, (2) computing
a set of possible values to calculate the watermark and then
selecting the optimum one with the help of an optimization scheme,
and (3) embedding the information-preserving optimum watermark into
EMR fields for right protection.

\begin{figure*}[htp]\scriptsize
\centering \includegraphics[angle=0,
width=1\columnwidth]{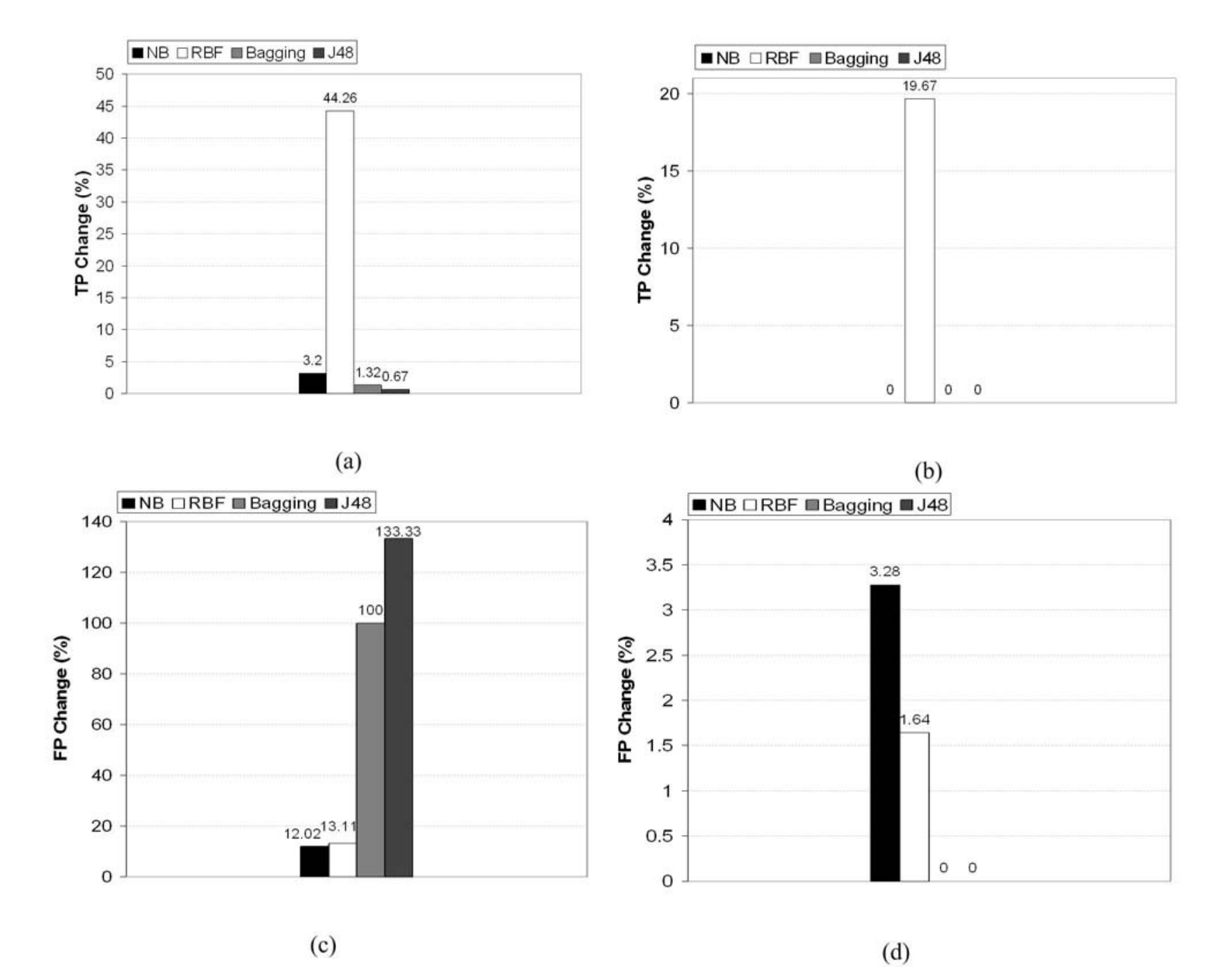} \caption{Effect of inserting
a watermark, using techniques that do not preserve information, in
attributes with different $C_{P}$ on various classification
statistics for diagnosing hypertension in the hypertension dataset.
(a) Change in TP(\%) when only one attribute with the highest
$C_{P}$ is watermarked. (b) Change in TP(\%) when only one attribute
with the lowest $C_{P}$ is watermarked. (c) Change in FP(\%) when
only one attribute with the highest $C_{P}$ is watermarked. (d)
Change in FP(\%) when only one attribute with the lowest $C_{P}$ is
watermarked.
}\label{fig:SelectiveAttributeWatermarkingClassificationStatistics}
\end{figure*}


For a quick reference, Table \ref{table:SymbolsTable} lists the
notations used in this paper. (We recommend that a reader should
refer to this table whenever he/she is in doubt about the
interpretation of an encountered symbol.)

\begin{table*}[t]
\scriptsize
  \centering
  \caption{Notations used in the paper.}
  \label{table:SymbolsTable}

\begin{tabular}{|l|l|l|l|}

\hline
\textbf{Symbol} & \textbf{Description} & \textbf{Symbol} & \textbf{Description}  \\
\hline
$EMR$ & Original database & $b$ & The watermark bit\\
\hline
$EMR_{W}$ & Watermarked database & $\mu$ & The mean of an attribute\\
\hline
\emph{R}& Total number of rows in a table (or dataset) & $r$ & A row in the database table\\
\hline
$F$ & The features set present in a table (or dataset) & $\sigma$ & The standard deviation of an attribute $a$\\
\hline
$TP$ & The number of  positive instances correctly detected & $min(a)$ & The minimum value of an attribute $a$\\
& as positive & & \\
\hline $FP$ & The number of  negative instances incorrectly detected
& $\varrho$ & The change in the value of an attribute in $EMR_{W}$\\
& as positive &  & after an attack \\
\hline
 $TN$& The number of  negative instances correctly detected & $max(a)$ & The maximum value of an attribute
 $a$\\
 & as negative & & \\
\hline
 $FN$& The number of  positive instances incorrectly detected & $C$ & The cardinality of $f$ \\
& as negative & & \\
\hline
 $C_{P}$& Classification potential of a feature in the database & $\hbar$ & Usability constraints matrix\\
 \hline
 $\Delta_{C_{P}}$& Change in classification potential of a feature in the &  $\eta_{f_{c}}$ & The exact value of watermark for a feature $f_{c}$ \\
 & database after changing its value & & row of the database relation\\
\hline
 $C_{S}$& Overall classification statistics of the database & $\Delta$ & A matrix containing secret parameters\\
\hline
 $C_{P_{O}}$& The classification potential of a feature in $EMR$ & $\tau$ & The watermark decoder used for watermark decoding\\
\hline
$C_{P_{W}}$& The classification potential of a feature in $EMR_{W}$ & $\gamma$ & A correction factor to adjust value of $\tau$\\
\hline
$a$& An attribute or column in the original database ($EMR$) & $EMR^{'}_{W}$ & A watermarked database after the malicious attacks\\
\hline
$x_{a}$& The value of an attribute $a$ & $l$ & The length of the watermark\\
\hline $IG(R,a)$& The information gain of an attribute $a$ in the
database & $\eta_{d_{f_{c}}}$ & Detected amount of change in
the value of a feature after \\
& & & an attack on the watermark bit \emph{b}\\
\hline $H(R)$& The entropy of $R$ rows & $\eta _{\Delta_{f_{c}}}$ &
The difference between the changes detected
in the value \\
& & & of a feature during the encoding and decoding process\\
\hline
$P$& The probability of a certain event & $W$ & Embedded watermark\\
\hline
$f_{c_{W}}$& The watermarked candidate feature $f_{c}$ & $W_{D}$ & Decoded watermark\\
\hline
$A$& Total number of attributes in the database & $EMR_{W_{P}}$ & $EMR$ watermarked by the proposed scheme\\
\hline
$m$& The number of features selected for watermarking & $ EMR_{W_{s}}$ & $EMR$ watermarked by Shehab's scheme\\
\hline

 $\beta _{f_{c}(min)}$ & Lower bound (percentage) for the watermark value for & $\Delta_{P}$ & The percentage difference between classification statistics \\
 & the feature $f_{c}$ & & when classifying $EMR$ and $EMR_{W_{P}}$\\
\hline $\beta _{f_{c}(max)}$& Upper bound (percentage) for the
watermark value for & $\Delta_{S}$ & The percentage difference
between classification
statistics \\

& the feature $f_{c}$ & & when classifying $EMR$ and $EMR_{W_{S}}$\\
\hline
$\beta $& An optimized value (percentage) of watermark for the & $DR$ & The percentage of positive instances correctly classified \\

 & feature $f_{c}$ & &as $TP$ \\
\hline
$V_{min}$& The normalized minimum value of an attribute in the database& $FAR$ & The percentage of negative instances incorrectly classified as $TP$\\

\hline $V_{max}$& The normalized maximum value of an attribute in
the database & $C_{P(BW)}$ & The classification potential of a
particular feature before\\

& & & watermarking\\
\hline $f$& The feature(s) set selected for watermarking &
$C_{P(AW_{P})}$ & The classification potential of a
particular feature after \\
& & & watermarking with our scheme\\
\hline $f_{c}$& A candidate feature to be watermarked &
$C_{P(AW_{S})}$ & The classification potential of a particular
feature after \\
& & & watermarking with the Shehab's scheme\\

\hline $p$& The number of particles in the swarm & $\Delta_{CP_{S}}$
& The difference in percentage between
$C_{P(AW_{S})}$ and \\
& & & $C_{P(BW)}$\\
\hline $g$& The maximum number of iterations for PSO &
$\Delta_{CP_{P}}$& The difference in percentage between
$C_{P(AW_{P})}$ and \\
& & &  $C_{P(BW)}$ \\

\hline

\end{tabular}
\end{table*}

It is worth mentioning here that -- for brevity -- in our pilot
studies, the reported results are on a Gynaecological database
collected by the authors of \cite{afridiog} from different tertiary
care hospitals. But our technique is not dependent on any particular
disease or any particular feature vector. Our technique takes the
features' set of any particular disease and calculates the
classification potential of all the features for classifying that
disease as depicted in section
\ref{sec:CandidateSubsetFeatureSelection}. Since the classification
potential of every feature, in the database of any disease, can be
calculated using equation (\ref{eq:CPFormula}); therefore, our
technique works for every disease with any set of features' vector.

\subsubsection{Ranking Features for Accurate Medical Diagnosis}
 \label{sec:CandidateSubsetFeatureSelection}
In order to rank features, we assign each feature in the feature
vector a value -- classification potential $(C_{P})$ -- that
reflects the importance of the feature in the diagnosis. The
classification potential $(C_{P})$  is then used to select features
(or attributes) that are to be watermarked. We rank features to
avoid a situation when a change in a feature will not only alter its
$(C_{P})$ but also overall classification statistics
($C_{S}$)\footnote{Classification statistics include TP, FP, TN, FN,
DR(\%), and FAR(\%) with $FAR = \frac{FP}{FP+TN}$ and $DR =
\frac{TP}{TP+FN}$.} of the diagnosis. These statistics are the
building blocks of the diagnosis rules; therefore, these statistics
must be preserved for preserving the diagnosis rules. As a
consequence, it is relevant to put the watermark in those features
that satisfy the following equation:

\begin{equation}\label{eq:CPDiff}   C_{P_{W}}-C_{P_{O}} = 0
\end{equation}
where $C_{P_{O}}$ and $C_{P_{W}}$ represent the classification
potential of a feature before and after insertion of a watermark
respectively. In order to do information-preserving watermarking, we
need to apply an optimization technique so that the above-mentioned
constraint is satisfied (the change in the value of high ranking
features are approximately zero). As a result, we preserve the
classification potential ($C_{P}$) and hence implicitly the overall
classification statistics ($C_{S}$). For ranking the features we use
a statistical parameter -- \emph{information gain}
\cite{mitchell1997machine}. The information gain, \emph{IG} for an
attribute \emph{a} may be calculated as:

\begin{equation}\label{eq:InfoGain}   IG(R,a)= H(R) - \sum_{j} P(a=v_{j})H(R|a=v_{j})
\end{equation}

with

\begin{equation}\label{eq:Entropy}   H(R)= -
\sum_{j \epsilon R}P(a=v_{j})log_{2}P(a=v_{j}),
\end{equation}

Where \emph{R} is the total number of rows in the table (or
dataset), $H(R)$ is the entropy of $R$ rows, $P$ represents the
probability, and $v_{j}$ is the value of an attribute \emph{a}. The
classification potential of an attribute \emph{a} is calculated
using the following equation:

\begin{equation}\label{eq:CPFormula}   C_{P_{a}}= \left ( \frac{IG(a)}{\sum_{n=1}^{A}IG(n)} \right )*100
\end{equation}

Where \emph{A} is the total number of attributes present in the
$EMR$. In order to prove our thesis that information-insensitive
(does not care about feature ranking) watermarking negatively
effects the classification accuracy, an empirical study is designed.
In this pilot study, we first sort the features on the basis of
their information gain. We then pick a feature from the top and the
bottom of the rank. We arbitrarily change the values of the selected
feature by inserting watermark and then analyze its impact on the
classification accuracy of well known classifiers (carefully chosen
from different learning paradigms): NaiveBayes, RBF Network, Bagging
and J48. The outcome of the empirical study is depicted in Figure
\ref{fig:SelectiveAttributeWatermarkingClassificationStatistics}.
Note that we ensure that the value of the features are never changed
by more than 5\% after inserting the watermark. Moreover, it is
ensured that usability constraints are also satisfied.

It is evident from Figure
\ref{fig:SelectiveAttributeWatermarkingClassificationStatistics}
that when only one top ranking feature is altered, the accuracies of
classifiers are significantly degraded (especially NB and RBF).
Hence it proves our thesis: \textit{In EMR, it is a must to do
information-preserving watermarking to reduce the rate of
misdiagnosis.} To conclude, after this step, we have identified top
ranking features in our features' vector. The next step is to
calculate the range of a watermark that does not significantly alter
the classification potential of each feature.

\subsubsection{Watermark Range Calculation}
 \label{WatermarkRangeCalculation}

We now need to develop a model for the range of values of (to be
constructed) watermark as a function of the classification potential
($C_{P}$) of the features.

Recall that the diagnosis rules are sensitive to the features with a
high classification potential. Therefore, a small change in high
ranking features is not acceptable because it deteriorates the
overall classification statistics $C_{S}$. To cater for this
situation, we calculate the lower ($\beta _{f_{c}(max)}$) and upper
bound ($\beta _{f_{c}(min)}$) of the watermark as a function of
$C_{P}$, $V_{min}$ and $V_{max}$ using equations (\ref{eq:BetaMin})
and (\ref{eq:BetaMax}) respectively for every selected feature
$f_{c}$ in the feature set $F$. We have empirically determined (See
Figure
\ref{fig:SelectiveAttributeWatermarkingClassificationStatistics})
that inverse relationship exists between the classification
potential of a feature and the amount of change it can tolerate,
leading us to equations (\ref{eq:BetaMin}) and (\ref{eq:BetaMax}).
According to these two equations, the attributes having greater
classification potential should be altered less to preserve the
classification statistics after the watermarking. In equations
(\ref{eq:BetaMin}) and (\ref{eq:BetaMax}), $1$ is added to the
denominator to avoid exceptions of infinity.  The second term in
equations (\ref{eq:BetaMin}) and (\ref{eq:BetaMax}) normalizes the
values of ($\beta _{f_{c}(max)}$) and ($\beta _{f_{c}(min)}$), where
$V_{min}$ and $V_{max}$ are the normalized minimum and maximum
values of a candidate attribute $f_{c}$, in the range (0,1).

\begin{equation}\label{eq:BetaMin} \beta _{f_{c}(min)}= \left ( \frac{1}{1+C_{P}} \right )*\left (\frac{V_{min}}{V_{max}+1}\right)
\end{equation}
\begin{equation}\label{eq:BetaMax} \beta _{f_{c}(max)}= \left ( \frac{1}{1+C_{P}} \right )*\left (\frac{V_{max}}{V_{min}+1}\right)
\end{equation}

\subsection{Watermark Creation}
 \label{sec:WatermarkCreation}

\subsubsection{Particle Swarm Optimization Algorithm}
 \label{ParticleSwarmOptimizationAlgorithm}

Particle swarm optimization (PSO) \cite{kennedy1995particle} is a
population based stochastic algorithm developed for continuous
optimization. The inspiration for PSO has come from the social
behavior of flocking birds. In PSO, each particle of a swarm
represents a potential solution. Each particle has its own set of
attributes including \emph{position, velocity } and a \emph{fitness
value} which is obtained by evaluating a fitness function at its
current position. The objective of particles is to search for a
global optimum. The algorithm starts with the initialization of
particles with random positions and velocities so that they can move
in the solution space. Then these particles search the solution
space for finding better solutions. Each particle keeps track of its
\emph{personal best} position found so far by storing the
coordinates in the solution space. The best position found so far by
any particle during any stage of the algorithm is also stored and is
termed as the \emph{global best} position. The velocity of every
particle is influenced by its personal best position
(autobiographical memory) and the global best position (publicized
knowledge). The new position for every particle is calculated by
adding its new velocity value to every component of its position
vector. The velocity $\upsilon_{i}$ of particle $i$ is updated using
the equation (\ref{eq:UpdateVelocity}) as follows:

\begin{equation}\label{eq:UpdateVelocity}
\begin{split}
 \upsilon_{ij}(t+1)& = \upsilon_{ij}(t) +
c_{1}*r_{1j}(t)*\left(y_{ij}(t)-x_{ij}(t)\right)\\
 & \quad + c_{2}*r_{2j}(t) *\left(y^{g}_{j}(t)-x_{ij}(t)\right)\\
\end{split}
\end{equation}

where $j$ is the index of the dimension of the problem, $t$
indicates a (unit) pseudo-time increment, $c_{1}$ and $c_{2}$ are
positive acceleration constants, $r_{1j}$ and $r_{2j}$ are random
values in the range $[0,1]$, $y_{ij}$ is the personal best position
of element $i$, $x_{ij}$ is the current position of the element $i$,
and $y^{g}_{ij}$ is the best position found by any particle of the
swarm. The position $x_{i}$ of the particle $i$ at any time interval
$t$ is added to its new velocity component to update its position
using the equation:
\begin{equation}\label{eq:UpdatePosition}
x_{i}(t+1) = x_{i}(t) + \upsilon_{i}(t+1)
\end{equation}
\subsubsection{Watermark Creation Algorithm}

We formulate the watermark creation process as a constrained
optimization problem. The objective function is formulated using
equation (\ref{eq:FitnessFunction}) as follows:

\begin{equation}\label{eq:FitnessFunction}
\begin{split}
& \forall f_{c}, \phantom{2} f_{c} \epsilon f, \phantom{2} \max
\beta_{f_{c}} \\
 & subject \phantom{1} to \\
 & 1. \phantom{1} \beta _{f_{c}(min)}\leq \beta_{f_{c}} \leq \beta _{f_{c}(max)} \\
 & 2.\phantom{1} CP_{W}= CP_{O}; \\
 & 3.\phantom{1} \frac{1}{\sigma \sqrt{2\pi }}e^{-\frac{1}{2}(\frac{x_{a}-\mu
}{\sigma})^{2}}  = \frac{1}{\sigma_{W} \sqrt{2\pi
}}e^{-\frac{1}{2}(\frac{x_{a_{W}}-\mu_{W} }{\sigma_{W}})^{2}}; \\
&  4.\phantom{1} min(a) = min(a_{W}); \\
&  5. \phantom{1} max(a) = max(a_{W}); \\
\end{split}
\end{equation}

Where $x_{a}$ represent the value of an attribute $a$ in the dataset
($EMR$) and $x_{a_{W}}$ is the value of an attribute $a_{W}$ in the
dataset ($EMR_W$), $\mu$ and $\sigma$ are the mean and standard
deviation of the attribute $a$ of dataset, and $\mu_{W}$ and
$\sigma_{W}$ are the mean and standard deviation of the attribute
$a_{W}$ in the watermarked dataset. The classification potentials --
$CP_{O}$ and $CP_{W}$ -- are calculated using equation
\ref{eq:CPFormula}. (Generally speaking, all usability constraints
should be satisfied during the process of watermarking.) The
motivation of above-mentioned usability constraints are: (1) the
distribution of data values before and after the watermark should
also remain the same (to be more precise, the probability density
function (pdf) of the original data should be preserved,
particularly for high ranking features), (2) the minimum and maximum
values of an attribute $a$ should remain the same before and after
watermarking i.e. $min(a) = min(a_{W})$ and $max(a) = max(a_{W})$,
(3) the constraints on the data type of any attribute should not be
violated, for example the number of children should not be changed
to floating point, and (4) the class to which an attribute belongs
prior to the watermarking should not be changed after inserting a
watermark. The classical optimization techniques are not suitable
because (1) they can get struck in local optima
\cite{renner2003genetic}, and (2) they cannot solve the problem in
realtime in case of large dimensional problems. It has been an
established fact that PSO is better suited for constrained
optimization problems compared with genetic algorithms, Memetic
algorithms (MAs), and Ant-colony optimization algorithms ( because
of its high success rate, better solution quality and less
processing time \cite{elbeltagi2005comparison},
\cite{hassan2005comparison}.

\begin{figure*}[htp]\scriptsize
\centering \includegraphics[angle=0,
width=1\columnwidth]{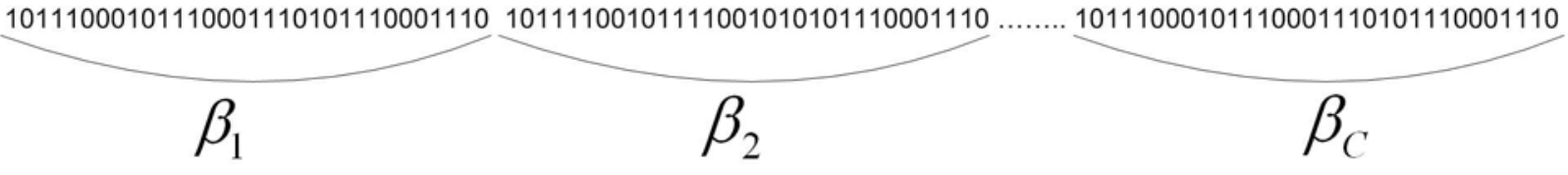} \caption{The structure
of the PSO particle.}\label{fig:ParticleStrucutre}
\end{figure*}

In our implementation of PSO, a particle consists of all $\beta$s.
Each $\beta$ is represented by a floating point number that denotes
the percentage change that the value of a feature can tolerate in
order to ensure that the usability constraints specified in the
usability constraints matrix $\hbar$ are met. As a result, a
particle's size is $32*C$, where C is the number of features. The
particle structure is depicted in Figure
\ref{fig:ParticleStrucutre}.

\begin{equation}\label{eq:ChangeAmount}
\eta_{(f_{c})} = \beta_{f_{c}}* f_{c}
\end{equation}

\begin{algorithm} [t]\label{alg:watermarkcreationAlgo}
\scriptsize \caption{WatermarkCreation}
\label{alg:watermarkcreationAlgo} \algsetup{indent=2em}

\begin{algorithmic}
\REQUIRE  $EMR$, $f$, Usability Constraints set $\hbar$ \ENSURE
\emph{Betamatrix} $\{\beta_{1},\beta{2},.....,\beta_{C}\}$

 \STATE Initialize swarm
$S_{W}$ of $p$ particles with positions and velocities in the range
$(\beta_{min}, \beta_{max})$ and $0\leq
\upsilon_{i}(0)\leq\upsilon_{max}(0)$ respectively

\STATE for every feature in the $f$ perform the following steps

\FOR {$iter=1$ to $iter_{max}$}

\FOR {$i=1$ to $p$}

\STATE note personal best and local best positions

\STATE Use swarm to calculate watermark to insert in $f$ subject to
$\hbar$

\STATE Evaluate the fitness of particle $i$ using equation
(\ref{eq:FitnessFunction})

\IF{\emph{stopping condition has met}}

 \STATE return $\{\beta_{1},\beta{2},.....,\beta_{C}\}$

\ELSE

\STATE Update personal best and global best positions

 \ENDIF

 \ENDFOR

 \ENDFOR
\STATE return $\{\beta_{1},\beta{2},.....,\beta_{C}\}$
\end{algorithmic}
\end{algorithm}

In the proposed scheme, the watermark value is calculated using the
optimized $\beta$ with the help of equation (\ref{eq:ChangeAmount}).
Once the optimum value of $\beta$ for each candidate attribute
$f_{c}$ is found, the algorithm is stopped.

\begin{figure*}[htp]\scriptsize
\centering \includegraphics[angle=0,
width=1\columnwidth]{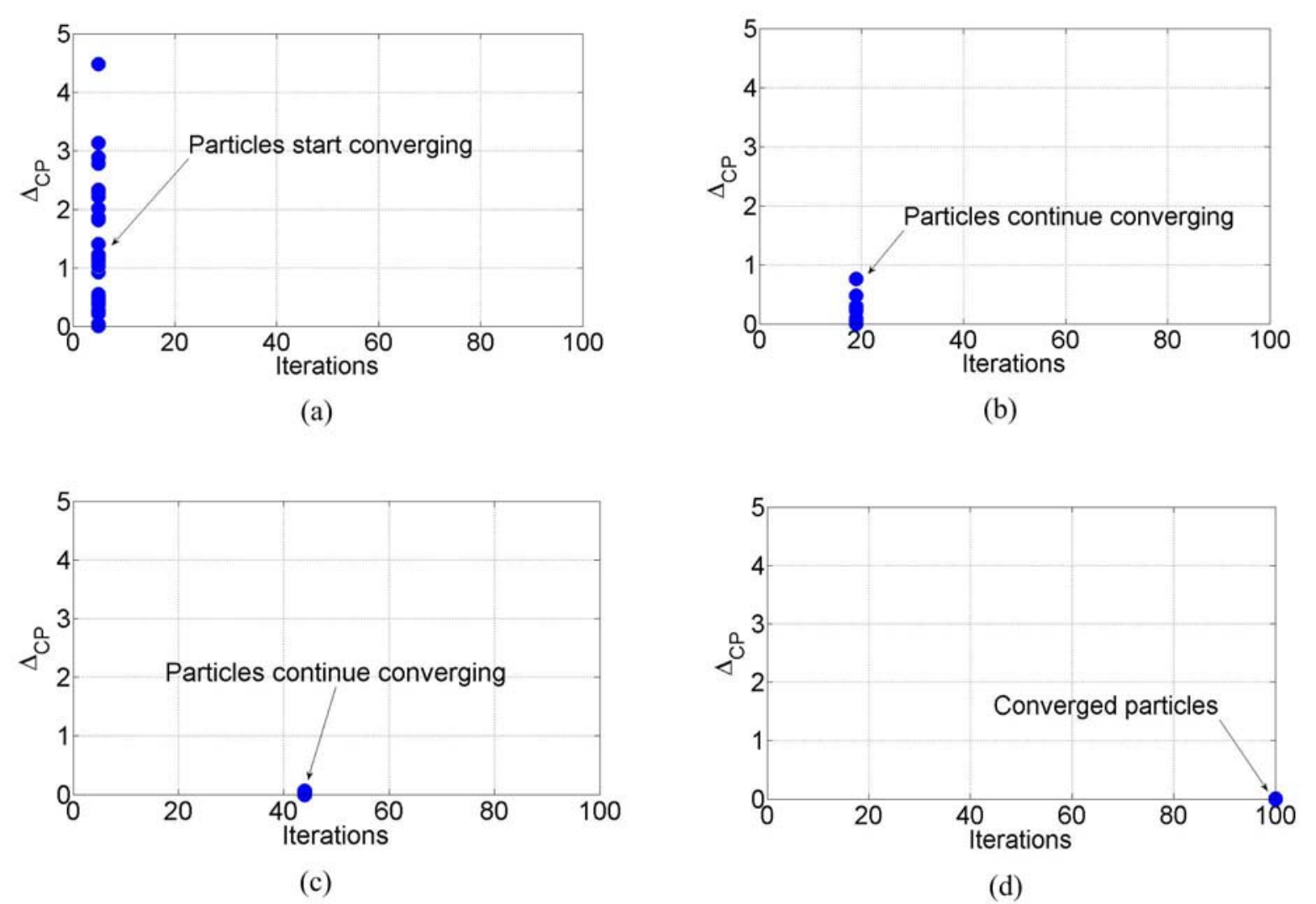} \caption{Motion of
particles to find optimum value for watermarking subject to
$\hbar$.}\label{fig:ParticlesMovement}
\end{figure*}

For brevity, we have only reported the motion of particles to find
the optimum value of $\beta$ -- satisfying the constraints ($\hbar$)
using the fitness function given in equation
(\ref{eq:FitnessFunction}) -- in Figure \ref{fig:ParticlesMovement}
for a candidate feature with the highest classification potential
$C_{P}$. It is evident from the figure that the particles do
converge towards an optimum value of a watermark with the progress
of PSO. Recall that the motivation of our work is to preserve the
classification potential ($C_{P}$) of high ranking features;
therefore, we plot $\Delta_{C_{P}}$ -- the difference between
$C_{P_{W}}$ and $C_{P_{O}}$ -- in Figure \ref{fig:ParticlesMovement}
to substantiate our argument.

The benefit of using PSO for calculating $\beta$ prior to watermark
insertion is that the process of optimization is not required during
the embedding process. As a result, the watermark can be inserted
online in realtime into every tuple of EMR. The important steps of
watermark creation algorithm are shown in Algorithm
\ref{alg:watermarkcreationAlgo}.

To ensure that the classification potential of a feature is not
altered once the watermark is inserted, EMR usability constraints
are specified by the set $\hbar$. If a particle violates the
usability constraints, its velocity is updated is such a way that it
quickly returns to the feasible space. A watermark consisting of $l$
bit strings is generated and the watermark value $\eta$ is added to
the value of the feature $f_{c}$ when a given bit is 0; otherwise,
its value is subtracted from the features' value. The algorithm
stops when the maximum number of iterations are done or $CP_{O}$ and
$CP_{W}$ have approximately equal values for a certain number of
iterations.

\subsection{Watermark Embedding}
 \label{sec:WatermarkEmbedding}

A bit string of length $l$ is generated as the watermark and the
best particle from PSO swarm is used to calculate and embed the
watermark value in relevant fields of EMR provided it satisfies the
usability constraints set $\hbar$. The watermark value $\eta$ is
calculated using the $\beta$ values in equation
(\ref{eq:ChangeAmount}). Since the length of the watermark is $l$;
therefore, the watermark value $\eta$ is calculated and inserted $l$
times in the $EMR$. The length \emph{l} of the watermark should be
carefully chosen because: (1) if it is too small, it makes the
watermark vulnerable to attacks, and (2) if it is too large, it
results in violating the usability constraints. Moreover, the
watermark should be created in realtime. After a number of empirical
studies, a length of $16$ bits is recommended that meets the
above-mentioned requirements. (It is important to emphasize that our
algorithm allows a user to chose a watermark length of 8, 16, 32 and
64 bits.)

\begin{algorithm}[t] \label{alg:watermarkembeddingAlgo}
\scriptsize
 \caption{EmbeddingWatermark}
\label{alg:watermarkembeddingAlgo} \algsetup{indent=2em}
\begin{algorithmic}
\REQUIRE   $EMR$, $Watermark W = {\beta_{1}, \beta_{2}, ... ,
\beta_{C}}$, $\hbar$

\ENSURE $EMR_{W}$, $\Delta$

\FOR{$q=1$ to $C$}

\FOR {$r=1$ to $R$}

\IF{$b_{i}==0$}

\STATE  $EMR_{W(r,q)} \Leftarrow EMR_{(r,q)} + \eta_{(r,q)}$ subject
to $\hbar$

\ENDIF
\STATE \emph{insert} $\eta_{(r,q)}$ \emph{into} $\Delta$

\IF{$b_{i}==1$}
\STATE  $EMR_{W(r,q)} \Leftarrow EMR_{(r,q)} - \eta_{(r,q)}$ subject
to $\hbar$

\ENDIF
\STATE \emph{insert} $\eta_{(r,q)}$ \emph{into} $\Delta$

 \ENDFOR

 \ENDFOR

\STATE \emph{return} $EMR_{W},\Delta$

\end{algorithmic}
\end{algorithm}

If the bit is 0 then the value of the selected feature in the matrix
$f$ is watermarked by adding $\eta$, which is calculated in the
watermark creation step, to its value. If the bit is 1, $\eta$ is
subtracted from the value of the selected feature. We use numeric
attributes to illustrate the watermark embedding procedure. The
watermark is created from Step 2 presented in section
\ref{sec:DeterminationofWatermarkParameters}. $\Delta$ is a matrix
that contains the statistics for each feature $f_{c}$, such as the
percentage change $\beta_{f_{c}}$, exact change $\eta$, and the
overall change. This matrix is continuously updated during the
watermarking embedding process to use it in the decoding stage. The
embedding algorithm is run $l$ times to embed the watermark multiple
times keeping in view the usability constraints $\hbar$. The
Algorithm \ref{alg:watermarkembeddingAlgo} lists the steps involved
in the watermark embedding stage.

\subsection{Watermark Decoding}
 \label{sec:WatermarkDecoding}

In the watermark decoding process, first step is to locate the
attributes which have been watermarked. To this end, we again rank
the attributes in $EMR_{W}$ using the procedure given in Section
\ref{sec:CandidateSubsetFeatureSelection}. We propose a novel
watermark decoder $\tau$, which calculates the amount of change in
the value of a feature that does not effect its classification
potential. The watermark decoder decodes the watermark by analyzing
one bit at a given time. The decoding phase consists of two steps:

Step 1. For every candidate feature $f_{c}$ of all the rows in
$EMR_{W}^{'}$, the watermark bits are detected starting from LSB
(least significant bit) and moving towards MSB (most significant
bit). The bits are detected in the reverse order compared with the
bits embedding order because it is easy to detect the effect of last
embedded bit of the watermark. This process is carried out using the
change matrix $\Delta$.

Step 2. The bits are then decoded using a majority voting scheme.

\subsection{Watermark Decoder}
 \label{sec:ResolutionOperator}
We now define a novel watermark decoder that takes into account the
degree of change, depending on its classification potential,
tolerated by each feature without violating the usability
constraints. It is defined as:
\begin{equation}\label{eq:ResolutionOperator}
\tau_{f_{c}} = \frac{\beta_{f_{c}}}{\gamma},
\end{equation}
where $ \gamma $ is the correction factor in the range (0, 1) and is
always greater than zero.

\begin{figure}[t]
\centering \includegraphics[angle=0,
width=0.60\columnwidth]{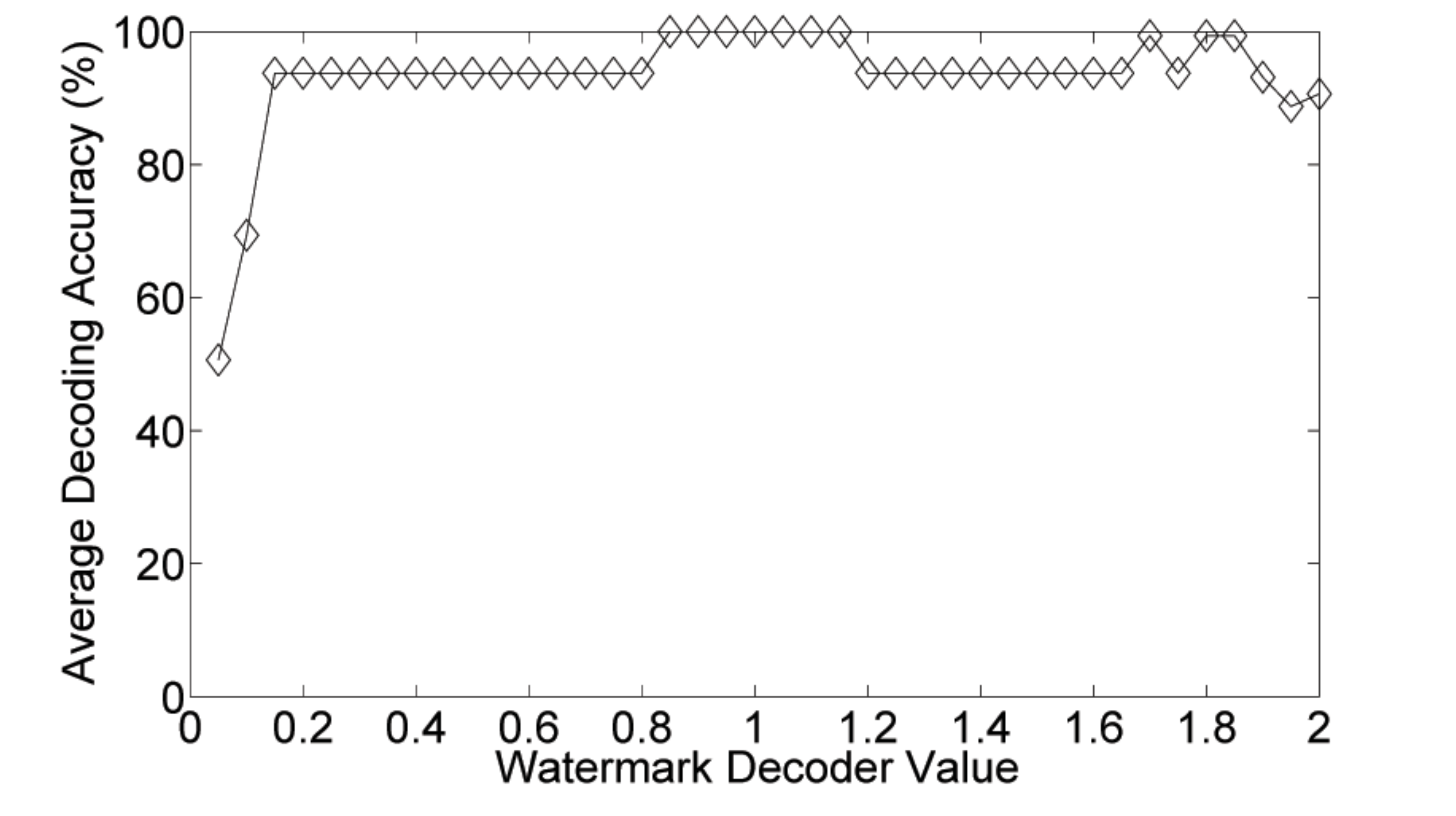}
\caption{Watermark decoding accuracy for different values of
$\tau$.}\label{fig:ResolutionOperatorandAccuracies}
\end{figure}

We have tested the operator with different values of $ \beta $ and $
\tau_{f_{c}} $ on different $EMRs$ as well. For brevity, we show
plot decoding accuracy of our technique on different combinations of
$\beta$
 and $\tau$ in Figure
\ref{fig:ResolutionOperatorandAccuracies} for only one selected
feature from the $EMR_{W}$. The decoding phase of our scheme is
dependent on the watermark decoder. The watermark decoder in turn
uses the knowledge of percentage change $\beta$ introduced during
the watermark embedding stage. As a consequence, our watermark
decoder also becomes information-preserving during calculation of
$\beta$; and hence decoding errors are reduced.

Our decoding algorithm selects one row at a time, for every
watermarked feature, and calculates $\eta_{d_{f_{c}}}$ using the
equation:
\begin{equation}\label{eq:percentChangeD}
\eta _{d_{f_{c}}} = \beta_{f_{c}}\ast(f_{c_{W}}),
\end{equation}
where $ \eta _{d(r,_{f_{c}})} $  is the detected amount of change in
the value of a feature after an attack on the watermark bit
\emph{b}. The algorithm computes the difference between
$\eta_{d_{f_{c}}}$ and $\eta_{_{f_{c}}}$ using the equation:

\begin{equation}\label{eq:percentChangeDiff}
\eta _{\Delta_{f_{c}}} = \eta_{d_{f_{c}}}- \eta_{f_{c}}
\end{equation}

Where $\eta _{\Delta_{f_{c}}}$ represents the difference between the
changes detected in the value of a feature during the decoding
process and the changes actually made in its value during the
encoding process. This value is compared with the watermark decoder
to correctly decode the bit. If $\eta _{\Delta_{f_{c}}}$ for a row
$r$ is less than or equal to zero then decoded bit is 1, and if the
value of $\eta _{\Delta_{f_{c}}}$ for that row lies between 0 and
$\tau$ then the decoded bit is 0. Recall, during the watermark
creation process the maximum value of $\beta$ is calculated subject
to the usability constraints $\hbar$. Similarly, $\beta$ is used to
compute the optimum value of watermark decoder $\tau$; therefore, if
for a row $r$ $\eta _{\Delta_{f_{c}}}$ is greater than $\tau$ then
the usability constraints must have been violated for that row $r$.
In this case, the algorithm decodes the watermark bit as $\times$.
For data to remain useful even after attacks, the number of bits
decoded as $\times$ will be very small for the whole dataset. As a
result, the majority voting used during the watermark decoding
process cancels the effects of such bits that ensures high watermark
decoding accuracy. After decoding a bit, the next bit is detected
using the same process. The watermark decoding steps are depicted in
Algorithm \ref{alg:WatermarkDecodingAlgo}.

\begin{algorithm} [t]\label{alg:watermarkdecoding}
\scriptsize
 \caption{WatermarkDecoding}
\label{alg:WatermarkDecodingAlgo} \algsetup{indent=2em}
\begin{algorithmic}

\REQUIRE  Altered EMR $EMR_{W}^{'}$, Data Change Matrix $\Delta$,
$f_{c}$, $l$

\ENSURE Detected Watermark $W_D$

\STATE $dt(1,2,...,l)\Leftarrow 0$ \STATE $Ones\Leftarrow 0$ \STATE
$Zeros\Leftarrow 0$

\FOR{$q=1$ to $C$}
 \FOR{$b=l$ to $1$}
 \FOR {$r=1$ to $R$}


\STATE Compute $\eta_{d_{q}}$ and $\eta_{\Delta_{q}}$ using
equations (\ref{eq:percentChangeD}) and (\ref{eq:percentChangeDiff})
respectively

\IF {$\eta_{\Delta_{q}} \leq 0$}
 \STATE $dt(r,b) \Leftarrow 1$
 \STATE $Ones=Ones+1$
  \ELSIF {$\eta_{\Delta_{q}} > 0$ and $\eta_{\Delta_{q}} \leq \tau$}
  \STATE $dt(r,b) \Leftarrow 0$
  \STATE $Zeros=Zeros+1$
  \ELSE [//usability constraints violated]

 \STATE $dt(r,b) \Leftarrow \times$

\ENDIF

\ENDFOR

 \ENDFOR
 \COMMENT {//Update the data}

\IF {$Ones>Zeros$}

\STATE $EMR^{'}_{W_{q}} = EMR^{'}_{W_{q}} + \eta_{d(q)}$

\ELSE

\STATE $EMR^{'}_{W_{q}} = EMR^{'}_{W_{q}} - \eta_{d(q)}$

 \ENDIF
\ENDFOR

\STATE $W_{D} \Leftarrow mode(dt(1,2,...,l)) $

\STATE return $W_{D}$
\end{algorithmic}
\end{algorithm}

The time complexity of our algorithm is $({l*C*R})$ where ${R}$ is
the total number of tuples in the database relation, $l$ is the
watermark length and $C$ is the number of selected subset features
from the relation in the database. Since $ l \ll R $ , and $ C \ll
R$, therefore, for large databases the time complexity of encoding
and decoding phases of proposed watermarking technique is $O(R)$. In the following subsection we give an example to highlight the working of our watermark embedding and watermark decoding algorithms.

\subsection{Example}
\subsubsection{Watermark Encoding}

Consider an EMR that has one table with three columns. The third column $D_{1}$ contains the value \emph{Yes} and \emph{No} indicating if a patient is suffering from the disease $D_{1}$  or not. Other two columns $A_{1}$ and $A_{2}$ contain numeric values for some vitals of the patient. In order to select the candidate feature for watermarking, we first calculate the classification potential of the features $A_{1}$ and $A_{2}$ using equation (\ref{eq:CPFormula}). After this step, we have identified $A_{1}$ as the candidate feature for watermarking because $A_{2}$, having the highest classification potential should not be watermarked. Suppose after applying the optimization scheme the optimum value of $\beta=1\%$ is found for watermark embedding. Now suppose we want to insert a 5-bits long watermark "11001" in the EMR. We calculate the value of $\eta$ to embed into $A_{1}$ using the equation:

\begin{equation*}\label{eq:Example1}
\eta = \beta. A_{1}
\end{equation*}

Our watermark embedding takes the most significant bit (MSB) of the watermark to embed it into $A_{1}$. For this purpose the algorithm works with one row at a time. Now, as the MSB of the watermark is 1, therefore, the new value of $A_{1}$ (we denote it by $A_{1_{W}}$ after embedding this bit will be:
\begin{equation*}\label{eq:Example2}
A_{1_{W}}  = A_{1}-\eta
\end{equation*}

Now, in order to embed second MSB our algorithm again applies the same mechanism, but the updated value of the attribute $A_{1}$ (which was $A_{1_{W}}$ ) is used for calculating new values of $\eta$ and $A_{1_{W}}$ . When our algorithm reaches at third MSB of the watermark then the new value of $A_{1_{W}}$  after embedding this bit would be:

\begin{equation*}\label{eq:Example3}
A_{1_{W}}  = A_{1}+\eta
\end{equation*}
The above mechanism is repeated until all rows of the EMR have been watermarked to generate $EMR_{W}$ . In $EMR_{W}$  both features $A_{1}$ and $A_{2}$ have the same classification potential which they had in EMR. The watermarking process is illustrated in Figure \ref{fig:WatermarkEmbeddingExample}.

\begin{figure}[htp]\scriptsize
\centering \includegraphics[angle=0,
width=0.35\columnwidth]{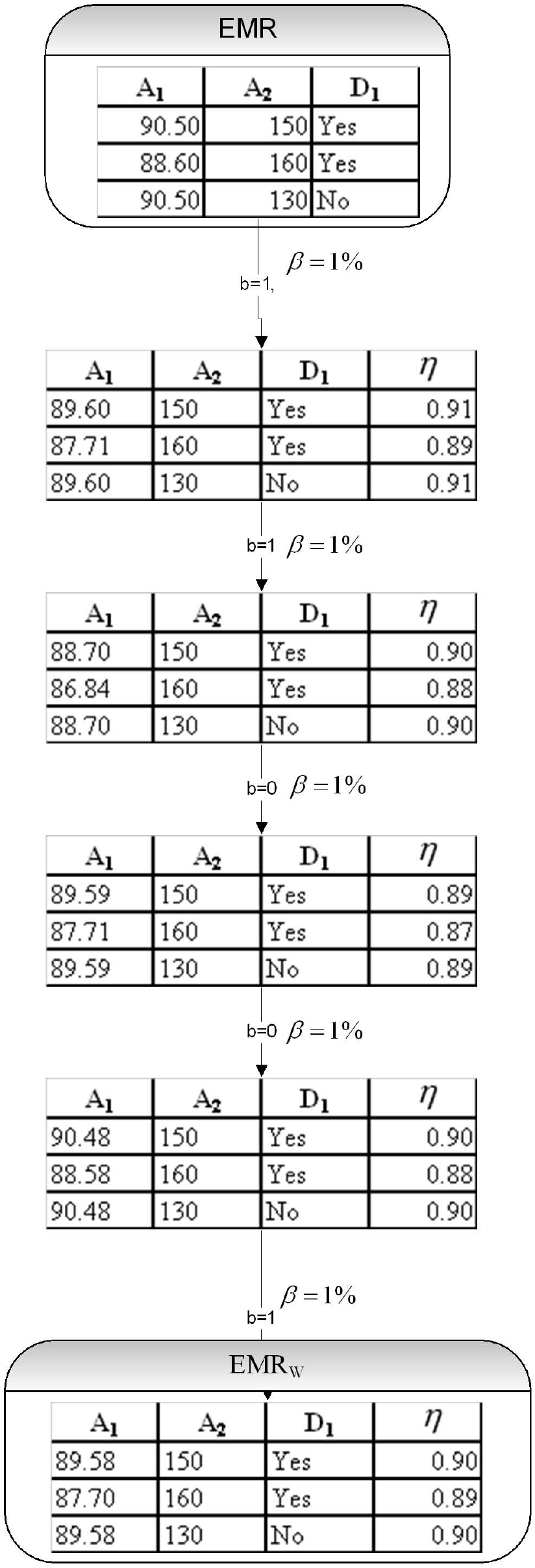} \caption{Watermark encoding process with $\beta = 1\%$ and watermark $W =11001$.}\label{fig:WatermarkEmbeddingExample}
\end{figure}
\subsubsection{Watermark Decoding}

In this phase, the features are again ranked based on their classification potential (calculated using equation (\ref{eq:CPFormula})). Now the feature with the least classification potential is $A_{1}$. So we have identified the watermarked feature. Now we use the values of $\beta$ (saved during the watermark encoding stage) to calculate the value of $\eta_{D}$ for $A_{1_{W}}$  as follows:

\begin{equation*}\label{eq:Example4}
\eta_{D} = \beta. A_{1_{W}}
\end{equation*}
We decode the embedded bits in reverse order in which they are encoded, that is, we start from the least significant bit (LSB) and move towards MSB. We use $\eta$ for embedding the last bit of the watermark and compute its difference from the $\eta_{D}$ using the equation:

\begin{equation*}\label{eq:Example5}
\eta_{\Delta}= \eta_{D} - \eta
\end{equation*}
We compare $\eta_{\Delta}$ with the value of our watermark decoding operator   to decode the last embedded watermark bit. After decoding the last embedded bit for all the rows, majority vote is taken and if the output of the majority voting process for that bit is 1 then 1 is saved as the detected bit and the values of feature $A_{1}$ in all the rows is updated using the equation:
\begin{equation*}\label{eq:Example6}
A_{1_{W}}  = A_{1_{W}} +\eta_{D}
\end{equation*}
And if the output of the majority voting is 0 then 0 is saved as the detected bit and the values of feature $A_{1}$ in all the rows is updated using the equation:
\begin{equation*}\label{eq:Example7}
A_{1_{W}}  = A_{1_{W}} -\eta_{D}
\end{equation*}

As mentioned earlier in this paper, the watermark decoding is done in the reverse order of the watermark embedding process: the last embedded bit is decoded first and so on. Therefore, in this example of watermark decoding (depicted in Figure \ref{fig:WatermarkDecodingExample}) the decoded watermark after applying the majority voting scheme on every row of the dataset at every step is 11001, which is same the embedded watermark.

\begin{figure}[htp]\scriptsize
\centering \includegraphics[angle=0,
width=0.45\columnwidth]{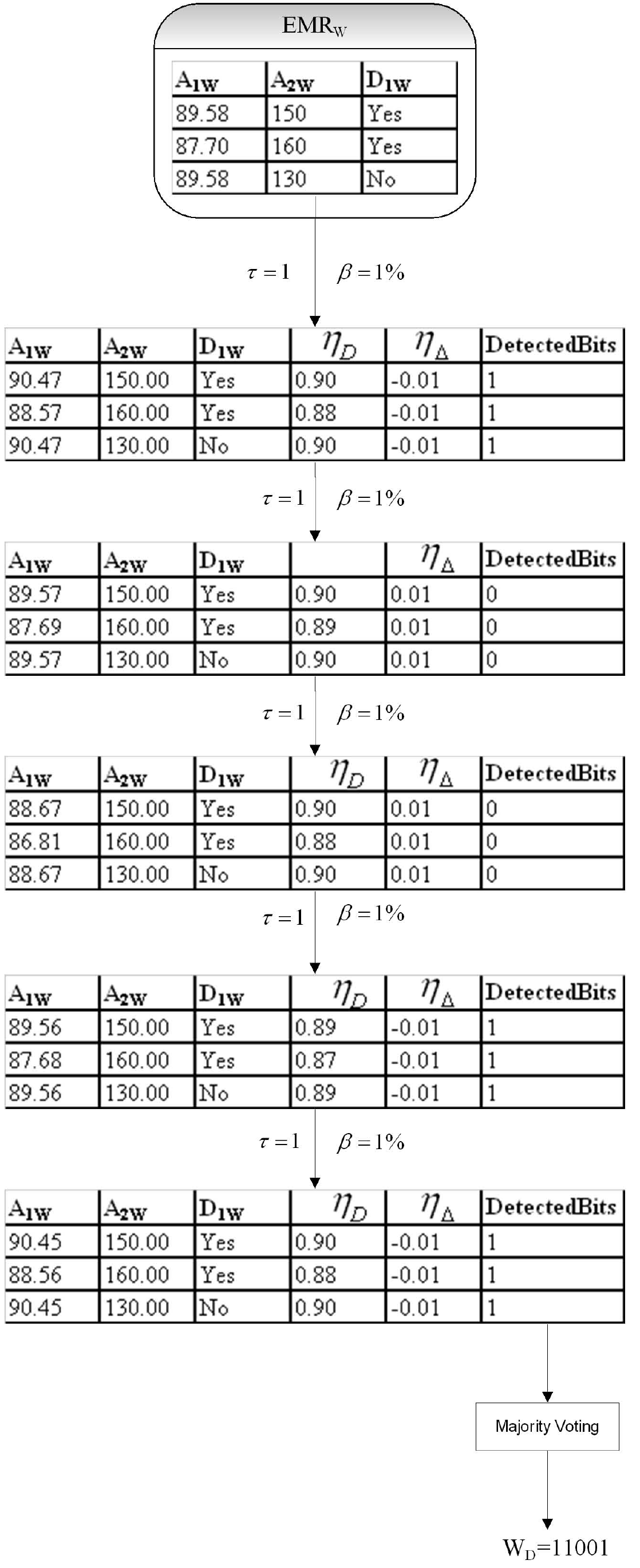} \caption{Watermark Decoding process with $\beta = 1\%$ and $\tau = 1$.}\label{fig:WatermarkEmbeddingExample}
\end{figure}

\section{Experiments and Results}
 \label{sec:ExperimentsandResults}

We have performed extensive experiments to test the effectiveness
and accuracy of the proposed watermarking technique. We have applied
our technique on two EMRs : (1) an oncology EMR provided for our
research by CureMD [http://www.curemd.com] and (2) a Gynaecological
EMR used by the authors of \cite{afridiog}.
For brevity, we report results on the second EMR that contains more
than 200,000 electronic health records. The usability constraints
specified in equation (\ref{eq:FitnessFunction}) are imposed during
watermark embedding. We also limit the maximum value of
$\beta_{f_{c}}$ to 2\%. We have performed our experiments on a
Microsoft SQL Server that is running on pentium IV computer with
1.73 GHz core 2 CPU with 1 GB of RAM.

It takes approximately $0.003$ seconds to embed a 16-bit watermark
in one feature of a health record. Similarly, it takes less than
$0.00026$ seconds per feature per record to decode it. The watermark
encoding takes relatively large time because the usability
constraints must be satisfied for every bit of the watermark in the
encoding process. The watermark decoder $\tau$ is computed using
equation (\ref{eq:ResolutionOperator}) to minimize the decoding
errors. (In the reported results we use \emph{p} = 100 particles,
\emph{g}=100, and \emph{l}=16.)

Recall that we give a choice to the owner of an EMR to chose a
watermark length by trading imperceptibility for efficiency. See in
Table \ref{table:WatermarkLengthsAndTime} to analyze the time
required to insert and detect varying length (8, 16, 32 and 64 bits)
watermarks. As a rule of thumb, we suggest using 16-bit watermark
because it provides very good imperceptibility with acceptable
processing overheads.

\begin{table}
[t]\scriptsize
  \centering
  \caption{Watermark creation, insertion and decoding time.}
  \label{table:WatermarkLengthsAndTime}

\begin{tabular}{|l|l|l|l|}

\hline
\textbf{$l$} & \textbf{Creation time} & \textbf{Insertion time} &\textbf{ Decoding time}\\
\hline
8-bits & 14 Milliseconds & 1.5 Milliseconds & 0.16 Milliseconds\\
\hline
16-bits & 27 Milliseconds & 3 Milliseconds & 0.26 Milliseconds\\
\hline
 32-bits & 44 Milliseconds & 7.6 Milliseconds & 0.38 Milliseconds\\
\hline
 64-bits & 66 Milliseconds & 15 Milliseconds & 3 Milliseconds\\
\hline
\end{tabular}
\end{table}
\subsection{Watermark Imperceptibility and Data Quality}
 \label{sec:WatermarkImperceptibilityandDataQuality}
\begin{figure*}[htpb]\centering
\subfigure[The normal probability density function (pdf) with normal
distribution for the attribute $SYSTOLIC\_BP$ in original data.]{%
{\includegraphics[width=0.60\columnwidth]{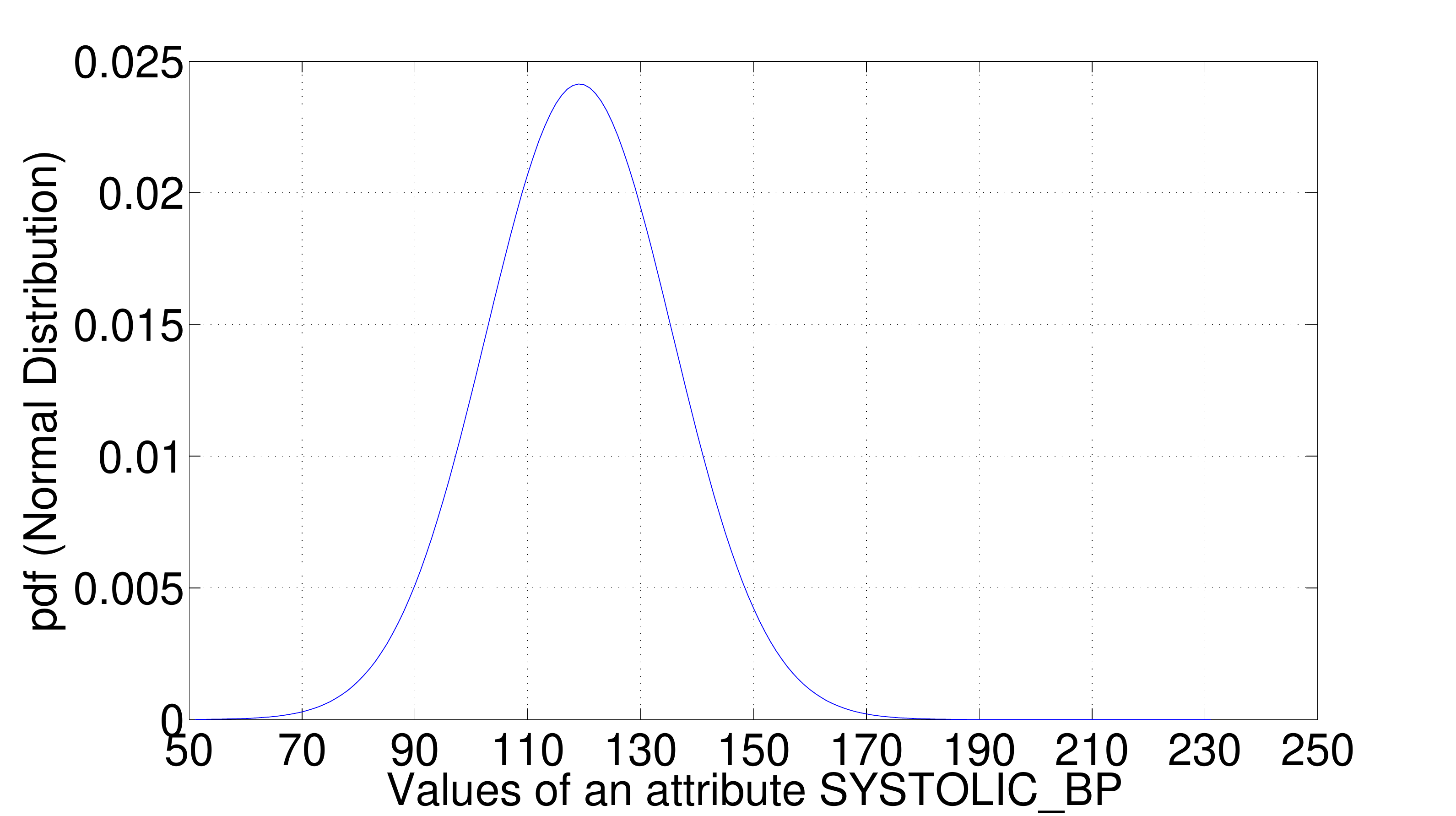}}}
\subfigure[The normal probability density function (pdf) with normal
distribution  for the attribute $SYSTOLIC\_BP_{W}$ in the
watermarked data.]{%
{\includegraphics[width=0.60\columnwidth]{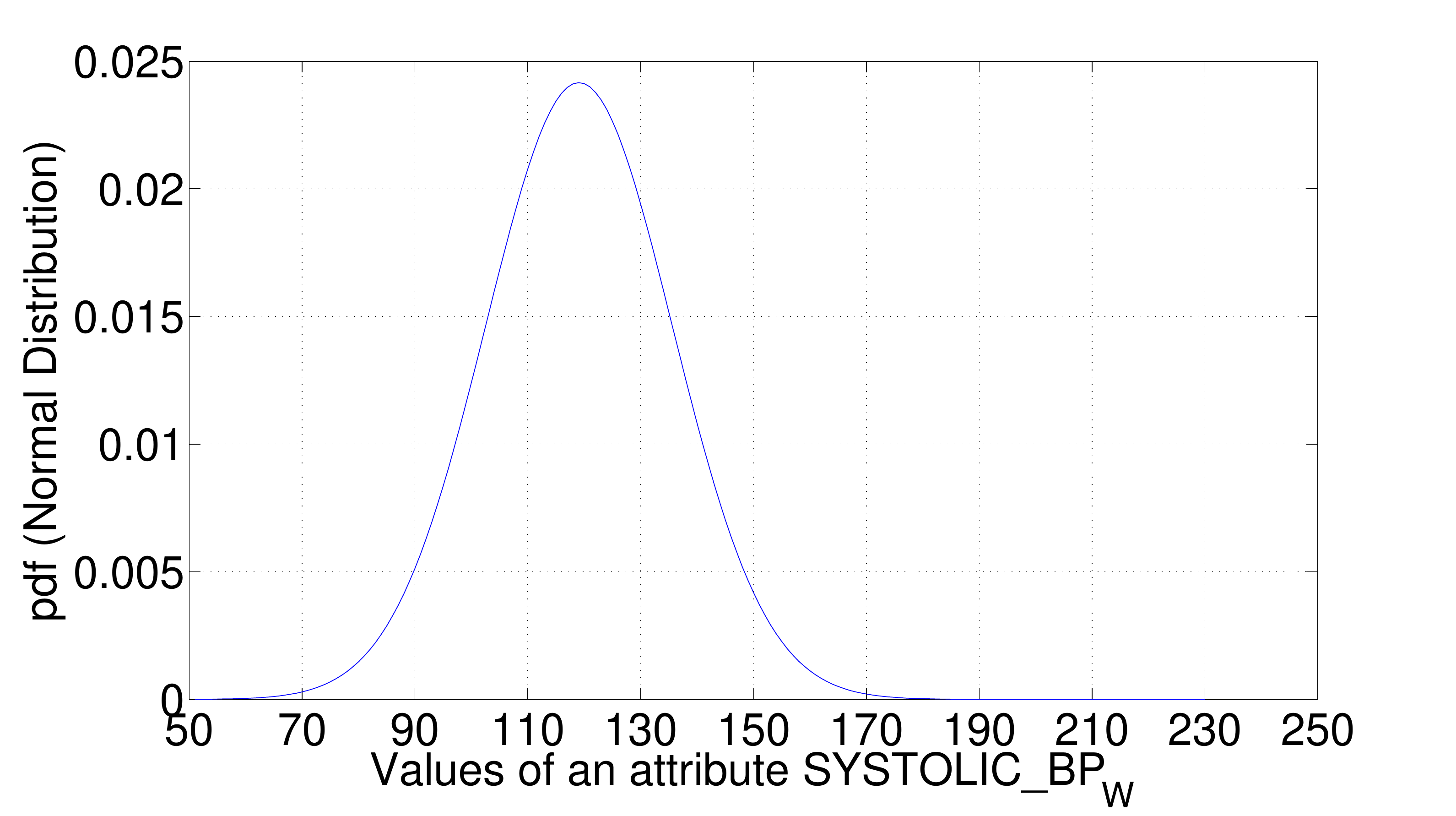}}}
\caption{The normal data distribution before and after watermarking
an attribute $SYSTOLIC\_BP$ in the \emph{EMR}. It is clear from the
figure that the distribution is preserved when the watermark was
embedded.} \label{fig:PDFs}
\end{figure*}

In this section, we show that once the $EMR$ is watermarked with the
proposed scheme, the watermark not only remains imperceptible but
also preserves the classification potential of all watermarked
features. We use two information theory measures -- Kullback -
Leibler Divergence ($D_{KL}$)\footnote{$D_{KL}(P \parallel
Q)=\sum_{i}P(i)log\frac{P(i)}{Q(i)}$.} and Jensen-Shannon Divergence
($JSD$)\footnote{$JSD(P \parallel Q)=\frac{1}{2}D(P \parallel
M)+\frac{1}{2}D(Q \parallel M)$,\\  \phantom{50} where
$M=\frac{1}{2}(P+Q)$.} to analyze the similarity between features'
distribution of watermarked and original datasets. The results are
tabulated in Table \ref{table:PDFComparison}. For brevity, we report
the comparison of pdfs for 2 features from the top and 2 from the
bottom of the ranked features, and also plot pdfs for the top ranking feature from both datasets in
Figure \ref{fig:PDFs}. We show the information-preserving
characteristic of our watermarking by classifying the watermarked
dataset with the classifiers that are used in Section
\ref{sec:CandidateSubsetFeatureSelection}. (We also add one more
rule-based classifier -- C4.5 -- to analyze the requirement of
preserving diagnosis rules.) We tabulate the classification
statistics in Table
\ref{table:ClassificationStatisticsAfterWatermarking} to show effect
of watermarking on them.

\begin{table}[t]

\scriptsize
  \centering
  \caption{The comparison between pdfs for $EMR$ and $EMR_{W}$.}
  \label{table:PDFComparison}
\begin{tabular}{|l|l|l|}

\hline
\multirow{2}{*}{\textbf{Features}} & \multicolumn{2}{|c|}{\textbf{Measure}}\\
\cline{2-3}
 & DKL & JSD\\
\hline
SYSTOLIC\_BP & 0 & 0\\
\hline
DIASTOLIC\_BP & 0 & 0\\
\hline
TEMPERATURE & 0.00001 & 0.000003\\
\hline
OPEN\_OS\_SIZE & 0.0002 & 0.00006\\
\hline
\end{tabular}
\end{table}

Remember that the motivation of proposing a information-preserving
watermarking scheme is to preserve the diagnosis of a medical
dataset. We have also included the scheme of Shehab
\cite{shehab2008watermarking} in our comparative study. It is
obvious from Table
\ref{table:ClassificationStatisticsAfterWatermarking} that though
the scheme of Shehab is imperceptible but it is not
information-preserving because it significantly alters the accuracy
of diagnosis. In comparison, our technique has preserved the
classification accuracy of different classifiers with the exception
of a classifier -- C4.5 -- where our scheme has improved $FAR$ by
reducing the $FP$. We believe this is due to the fact that the
proposed scheme considers the correlation present among
classification potential of all the candidate attributes for
watermarking along with the data usability constraints. In
comparison, the technique of Shehab only considers the last three
usability constraints given in equation (\ref{eq:FitnessFunction})
and does not take into account the classification potential of a
candidate attribute and its correlation with the decision.

\begin{table*}
[t]\scriptsize
  \centering
  \caption{Effect of different watermarking schemes on various classification statistics.}
  \label{table:ClassificationStatisticsAfterWatermarking}
\begin{tabular}{|l|l|l|l|l|l|l|}
\hline
\textbf{Classifier} & \textbf{Measure} & \textbf{$EMR$} & \textbf{$EMR_{W_{P}}$} & \textbf{$EMR_{W_{S}}$} & \textbf{$|\Delta_{P}|$} & \textbf{$|\Delta_{S}|$}\\
 \hline
 \multirow{5}{*}{\textbf{Bagging}} & TP & 152 & 152 & 124 & 0 & 18.52\\
\cline{2-7}
 & FP & 5 & 5 & 5 & 0 & 0\\
\cline{2-7}
 & TN & 3569 & 3569 & 3569 & 0 & 0\\
\cline{2-7}
 & \textbf{FN} & 30 & 30 & 58 & 0 & \textbf{92.51}\\
 \cline{2-7}
 & DR(\%) & 83.52 & 83.52 & 68.13 & 0 & 18.43\\
\cline{2-7}
 & FAR(\%) & 0.14 & 0.14 & 0.14 & 0 & 0\\
\hline
\multirow{5}{*}{\textbf{J48}} & TP & 150 & 151 & 137 & 0.75 & 8.52\\
\cline{2-7}
 & \textbf{FP} & 3 & 3 & 6 & 0 & \textbf{100}\\
\cline{2-7}
 & TN & 3571 & 3571 & 3568 & 0.01 & 0.1\\
\cline{2-7}
 & FN & 32 & 31 & 45 & 2.35 & 41.22\\
 \cline{2-7}
 & DR(\%) & 82.34 & 82.96 & 75.33 & 0.75 & 8.51\\
\cline{2-7}
 & FAR(\%) & 0.08 & 0.08 & 0.17 & 0 & 112.5\\
 \hline
\multirow{5}{*}{\textbf{C4.5}} & \textbf{TP} & 150 & 151 & 122 & 0.67 & \textbf{18.67}\\
\cline{2-7}
 & FP & 3 & 2 & 2 & 33.33 & 33.33\\
\cline{2-7}
 & TN & 3571 & 3571 & 3572 & 0.01 & 0.03\\
\cline{2-7}
 & FN & 32& 33 & 60 & 3.13 & 87.5\\
 \cline{2-7}
 & \textbf{DR(\%)} & 82.42 & 81.87 & 67.03 & 0.75 & \textbf{8.51}\\
\cline{2-7}
 & FAR(\%) & 0.08 & 0.06 & 0.06 & 25 & 25\\
 \hline
\multirow{5}{*}{\textbf{RBF}} & TP & 61 & 61 & 64 & 0 & 4.96 \\
\cline{2-7}
 & FP & 20 & 19 & 11 & 3.77 & 45.29\\
\cline{2-7}
 & TN & 3554 & 3555 & 3563 &  0.02 & 0.25\\
\cline{2-7}
 & FN & 121 & 121 & 118 & 0 & 2.48\\
 \cline{2-7}
 & DR(\%) & 33.43 & 33.43 & 35.08 & 0 & 4.94\\
\cline{2-7}
 & \textbf{FAR(\%)} & 0.56 & 0.54 & 0.3 & 3.57 & \textbf{46.43}\\
 \hline
\multirow{5}{*}{\textbf{NB}} & TP & 125 & 125 & 126 & 0 & 0.6\\
\cline{2-7}
 & FP & 183 & 187 & 199 & 2.26 & 8.83\\
\cline{2-7}
 & \textbf{TN} & 3391& 3387 & 3375 & 0.11 & \textbf{0.47}\\
\cline{2-7}
 & FN & 57 & 57 & 56 & 0 & 1.97\\
 \cline{2-7}
 & DR(\%) & 68.72 & 68.72 & 69.13 & 0 & 0.6\\
\cline{2-7}
 & FAR(\%) & 5.12 & 5.23 & 5.57 & 2.15& 8.79\\
 \hline
\end{tabular}

\end{table*}

\subsection{Preserving Classification Potential of High Ranking Features}
 \label{sec:ClassificationPotentialPreservation}

Our pilot studies reveal, if an $EMR$ is watermarked without the
knowledge of classification potential $C_{P}$ of a feature, the
diagnosis rules are significantly changed; as a consequence, the
misdiagnosis increases. This is a serious issue because if a patient
has a disease $A$ and after watermarking of features, the patient is
diagnosed for disease $B$. If the treatment of disease $B$ is $100$
times more expensive, then the patient ends up not only receiving
expensive but also incorrect treatment. Therefore, a watermarking
scheme has to be information-preserving to avoid such scenarios.

In Table \ref{table:PerformanceComparisonCP}, we show the effect of
watermarking on classification potential $C_{P}$. In the table,
$C_{P(BW)}$ is the classification potential of a particular feature
before watermarking and $C_{P(AW)}$ is the classification potential
of that feature after watermarking. It is obvious from Table
\ref{table:PerformanceComparisonCP} that information-preserving
watermarking scheme is able to preserve the classification potential
of a feature after watermarking; as a result, a patient is still
diagnosed accurately. In comparison, the threshold based technique
of Shehab alters $C_{P}$ of several features; as a result, the
overall classification statistics $C_{S}$ are significantly changed.

\begin{table*}
[t]\scriptsize
  \centering
  \caption{Effect of watermarking of $EMR$ on $C_{P}s$ of various features.}
  \label{table:PerformanceComparisonCP}
\begin{tabular}{|l|l|l|l|l|l|}

\hline
\textbf{Feature Name} & \textbf{$C_{P(BW)}$} & \textbf{$C_{P(AW_{P})}$} & \textbf{$C_{P(AW_{S})}$} & \textbf{$|\Delta_{CP_{P}}|$} & \textbf{$|\Delta_{CP_{S}}|$}\\
\hline
SYSTOLIC\_BP & 17.25 & 17.25 & 22.55 & 0 & 30.72 \\
\hline
DIASTOLIC\_BP & 15.24 & 15.24 & 20.28 & 0 & 33.07\\
\hline
USD\_OBSTETRICAL\_BPD & 10.53 & 10.53 & 10.91 &  0 & 3.61\\
\hline
RANDOM\_BLOOD\_SUGAR\_LEVELS & 7.58 & 7.58 & 3.92 & 0 & 48.28\\
\hline
USD\_OBSTETRICAL\_FL & 7.43 & 7.43 & 8.51 & 0 & 14.54\\
\hline
GRAVIDA & 6.39 & 6.39 & 8.52 & 0 & 33.33\\
\hline
AGE & 5.4 & 5.4 & 1.81 & 0 & 66.48\\
\hline
PULSE\_RATE & 4.28 & 4.28 & 0.89 & 0 & 79.21\\
\hline
HEIGHT & 4.1 & 4.1 & 0 & 0 & 100\\
\hline
PARITYB424MONTHS & 3.11 & 3.11 & 4.14 & 0 & 33.12\\
\hline
RESPIRATORY\_RATE & 2.94 & 2.94 & 3.92 & 0 & 33.33\\
\hline
WEIGHT & 2.88 & 2.88 & 1 & 0 & 65.28\\
\hline
PARITYAFTER24MONTHS & 2.43 & 2.43 & 3.23 & 0 & 32.92\\
\hline
RENAL\_FUNCTION\_UREA & 2.09 & 2.09 & 1.63 & 0 & 22.01\\
\hline
FUNDAL\_HEIGHT & 2.06 & 2.06 & 1.72 & 0 & 16.5\\
\hline
PULSE & 1.34 & 1.34 & 1.64 & 0 & 22.39\\
\hline
LFT\_ALT & 0.83 & 0.83 & 1.1 & 0 & 32.53\\
\hline
MENARCHE & 0.74 & 0.74 & 0.99 & 0 & 33.78\\
\hline
TLC & 0.63 & 0.63 & 0.89 & 0 & 41.27\\
\hline
RENAL\_FUNCTION\_CREATININE & 0.6 & 0.6 & 0.8 & 0 & 33.33\\
\hline
STATION & 0.6 & 0.6 & 0.79 & 0 & 31.67\\
\hline
PLATELET\_COUNT & 0.4 & 0.4 & 0 & 0 & 100\\
\hline
CF\_FUNDAL\_HEIGHT & 0.38 & 0.38 & 0 & 0 & 100\\
\hline
APPROX\_FETAL\_WEIGHT & 0.27 & 0.27 & 0.36 & 0 & 33.33\\
\hline
TEMPERATURE & 0.22 & 0.22 & 0 & 0 & 100\\
\hline
OPEN\_OS\_SIZE & 0.2 & 0.2 & 0.26 & 0 & 30\\
\hline

\end{tabular}
\end{table*}

Preserving the classification statistics of $EMR$, which in turn
preserves the diagnosis rules, is the most important requirement
that the proposed technique meets. For brevity, we only show a
subset of diagnosis rules for hypertension that are discovered by
J48 classifier for the hypertension dataset in Table
\ref{table:ClassificaionRulesPreserved}. It is easy to conclude that
the diagnosis rules do not change before and after watermarking
using our technique. In contrast, the technique of Shehab alters the
classification rules. See the additional terms appearing in the
diagnosis rules of J48 classifier once it is operating on the
dataset that is watermarked with Shehab's technique. (Remember that
the technique of Shehab is not information-preserving because it
does not take into account the impact of high ranking features on
the diagnosis.) This further validates our idea of ranking features
on the basis of classification potential before calculating the
watermark.

\begin{table*}
[t]\scriptsize
 \caption{Effect of different watermarking techniques on classification rules.}
  \centering
  \label{table:ClassificaionRulesPreserved}
\begin{tabular}{|l|l|l|}

\hline
\textbf{$EMR$} & \textbf{$EMR_{W_{P}}$} & \textbf{$EMR_{W_{S}}$}\\
\hline If SYSTOLIC\_BP $<=$ 120 And GRAVIDA $<=$ 1& If SYSTOLIC\_BP $<=$ 120 And GRAVIDA $<=$ 1 & If SYSTOLIC\_BP $<=$ 122 And GRAVIDA $<=$ 1\\
And APPROX\_FETAL\_WEIGHT $>$ 1.5  And &  And APPROX\_FETAL\_WEIGHT $>$ 1.5 And & And APPROX\_FETAL\_WEIGHT $>$ 1.5  And\\
 STATION $>$ 0  And OPEN\_OS\_SIZE $<= $3.5 & And STATION $>$ 0  And OPEN\_OS\_SIZE $<= $3.5& And STATION $>$ 0 And OPEN\_OS\_SIZE $<= $3.5\\
  And DIASTOLIC\_BP $<=$ 70   Then & And DIASTOLIC\_BP $<=$ 70   Then & \textbf{And HEIGHT $>$ 160.5}  Then \\
  Hypertension = Yes &Hypertension = Yes &Hypertension = Yes  \\
\hline If SYSTOLIC\_BP $<=$ 120 And GRAVIDA $<=$ 1& If SYSTOLIC\_BP $<=$ 120 And GRAVIDA $<=$ 1 & If SYSTOLIC\_BP $<=$ 122 And GRAVIDA $<=$ 1\\
And APPROX\_FETAL\_WEIGHT $>$ 1.5  And &  And APPROX\_FETAL\_WEIGHT $>$ 1.5 And & And APPROX\_FETAL\_WEIGHT $>$ 1.5  And\\
 STATION $>$ 0  And OPEN\_OS\_SIZE $<= $3.5 & And STATION $>$ 0  And OPEN\_OS\_SIZE $<= $3.5& And STATION $>$ 0 And OPEN\_OS\_SIZE $<= $3.5\\
And DIASTOLIC\_BP $>$ 70  Then & And DIASTOLIC\_BP $>$ 70 Then & \textbf{And HEIGHT $<=$ 160.5} Then \\
   Hypertension = No & Hypertension = No & Hypertension = No  \\
\hline If SYSTOLIC\_BP $<=$ 120 And GRAVIDA $<=$ 1& If SYSTOLIC\_BP $<=$ 120 And GRAVIDA $<=$ 1 & If SYSTOLIC\_BP $<=$ 122 And GRAVIDA $<=$ 1\\

And PULSE\_RATE $>$ 76 Then  &  And PULSE\_RATE $>$ 76  Then  & And PULSE\_RATE $<=$ 76  And \textbf{PULSE\_RATE $>$ 74}\\
Hypertension = No & Hypertension = No &  \textbf{And AGE $>$ 26 And AGE $<=$ 32} Then\\
& & Hypertension = No  \\
\hline
If SYSTOLIC\_BP $<=$ 160 And SYSTOLIC\_BP $>$ 120 & If SYSTOLIC\_BP $<=$ 160 And SYSTOLIC\_BP $>$ 120& If SYSTOLIC\_BP $<=$ 162 And SYSTOLIC\_BP $>$ 122\\
And HEIGHT $>$ 160.5  And STATION $>$ 2 And & And HEIGHT $>$ 160.5  STATION $>$ 2 And & And HEIGHT $>$ 160.5  And STATION $>$ 2 And \\
PULSE\_RATE $>$ 87 Then & PULSE\_RATE $>$ 87  Then & \textbf{HAEMOGLOBIN\_LEVELS $<=$ 8} Then \\
Hypertension = Yes & Hypertension = Yes & Hypertension = Yes \\
\hline

\hline
\end{tabular}

\end{table*}

Moreover, it is equally important to demonstrate that a watermarking
technique is also resilient to malicious attacks. During the
robustness study, only the proposed technique is evaluated because
techniques like Shehab's do not meet the first requirement because
they do not account for the classification potential of any
attribute before watermarking it. Moreover they do not give any
mechanism to decide which attribute is most suitable for
watermarking without disturbing the classification potential of the
dataset before and after watermarking. On the other hand, in our
technique we first examine the classification potential of each
candidate attribute to be watermarked and then the watermark range
for each attribute to be watermarked is calculated using equations
(\ref{eq:BetaMin}) and (\ref{eq:BetaMax}).

\subsection{Resilience to Various Attacks}
 \label{sec:ResiliencetoVariousAttacks}
In this study, different types of attacks are generated on the
watermarked data. A watermarking technique must meet two
requirements: (1) an intruder is not able to locate the watermark,
and (2) an intruder is not able to corrupt the watermark without
significantly degrading the quality of data.

Suppose Alice is the owner of an $EMR$ and she embeds a watermark
$W$ in the electronic records; as a result, the watermarked EMR is
$EMR_{W}$. Alice now wants to share $EMR_{W}$ with Bob. In the
meantime, Mallory (an intruder) wants to remove the watermark from
the dataset so that Alice is unable to claim ownership of its EMR in
a court of law.

The robustness study is conducted with the following assumptions:
(1) Mallory does not have access to the original $EMR$, and (2)
Mallory does not have the resources to acquire secret information
like candidate subset $f_{c}$ data of $EMR$, overall data change
$\Delta$, usability constraints $\hbar$, change in a particular
feature $\beta_{f_{c}}$, watermark length \emph{l}, and the
watermark decoder $\tau_{f_{c}}$. These assumptions (though
realistic) make the task of Mallory relatively difficult because he
has to corrupt the watermark without compromising the usability of
the medical data. Mallory may calculate the classification potential
of features in the watermarked dataset but he has to face a dilemma
of preserving the diagnostic rules of the dataset while trying to
corrupt the watermark.

Therefore, in a worst case scenario, if he is able to successfully
alter the watermarked feature while preserving usability constraints
(which he is not aware of), even then our technique is able to
successfully extract the watermark because we insert the watermark
in each row of the database and an attack on a set of rows does not
significantly affect the decoding process to a great extent. Also,
to deceive the attacker, apart from low ranking features, we select
some high ranking (except top $t$ features with respect to
classification potential, where $t$ can be decided by the data
owner) features for watermarking with classification potential
within a certain threshold such that watermarking them does not
alter the classification potential of any feature. And we have used
the majority voting scheme that significantly reduces the
probability of incorrect decoding on the basis of attacking few rows
only. Since the attacker has no knowledge of original data and the
embedded watermark, therefore, to corrupt the watermark -- he may
make random changes in the watermark dataset. So for him, at every
trial, the probability of success ($p$) is 0.5 for successfully
deleting one watermark bit from one row of the watermarked dataset.
But we embed watermark in each row of the dataset and take majority
voting while decoding the embedded watermark bit, so attacker has to
successfully attack at least half the total number of rows $R$ in
the dataset. And attack on one row is independent of the attack on
other rows, so according to the multiplication rule of the
probability, the probability of successfully attacking $\frac{R}{2}$
rows is $(0.5)^{\frac{R}{2}}$. But practically speaking, EMRs have
large number of health records and for such EMRs the probability of
successful attack even on a single watermark bit is approximately
zero.

successfully attacking a watermark bit is virtually zero.

\emph{Therefore, the knowledge of classification potentials of the
features will not be useful for him to corrupt the watermark unless
he deteriorates the data by violating the usability constraints.}
This model is validated by the experimental results, that proves
resilience of our technique against all types of malicious attacks.

In the robustness study, three attacks are studied: (1) Insertion,
(2) Deletion, and (3) Alteration.
\subsection{Insertion Attacks}
 \label{sec:InsertionAttacksResults}
\begin{figure*}[htp]\scriptsize
\centering \includegraphics[angle=0,
width=1\columnwidth]{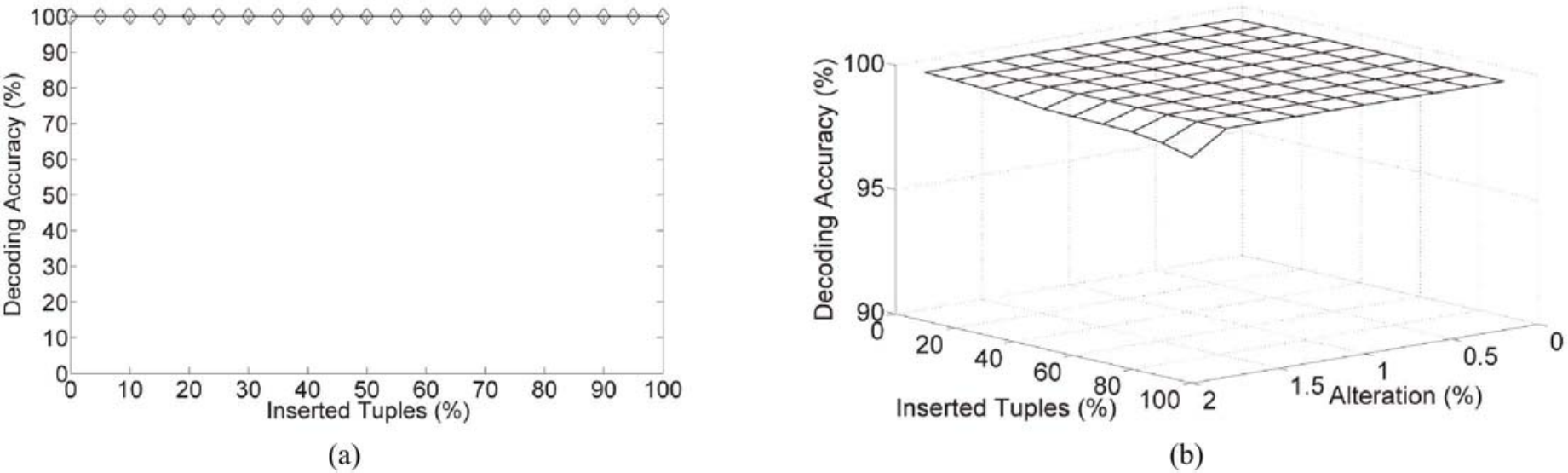} \caption{Resilience
of proposed technique against various insertion attacks. (a)
Resilience to $ \alpha$-duplicate insertion attacks. (b) Resilience
to $( \alpha,\varrho $) insertion attacks.
}\label{fig:InsertionAttacks}
\end{figure*}

Since we embed watermark in each row, and we do not target the order
of tuples during watermark embedding; therefore, insertion of new
tuples does not disturb the embedded watermark. In the decoding
stage, to achieve high detecting accuracy we take the majority
voting over all the rows for detecting the embedded watermark.
Moreover, we do not use any marker tuples for watermark embedding
hence inserting new rows does not make our scheme vulnerable to
synchronization errors (which might occur if insertion of new
records perturbs the order of marker tuples.)
 For insertion attacks,
Mallory can utilize two mechanisms to insert new tuples into EMR. He
might insert $\alpha $ new records by replicating the original
records. In this way, $\alpha$-duplicate records are inserted into
EMR. It is clear from Figure \ref{fig:InsertionAttacks} that the
proposed technique is resilient to $\alpha$-duplicate insertion
attack because it does not target any specific tuple(s) for
watermark embedding.

The second possibility is to randomly generate fake records and
insert them into the EMR. Since Mallory has no information about the
original EMR; therefore, he will generate useless records. He may
insert new records with the values within the range
($\mu_{W}\pm\varrho\sigma_{W}$) for a specific feature. This type of
attack is called ($ \alpha, \varrho$)- insertion attacks. It is
obvious from Figure \ref{fig:InsertionAttacks} that the proposed
technique is resilient (generally speaking) to ($\alpha, \varrho $)-
insertion attacks and we believe this is because it does not target
any specific tuple to insert the watermark.
\subsection{Deletion Attacks} \label{sec:DeletionAttacksResults}
\begin{figure}[htp]\scriptsize
\centering \includegraphics[angle=0,
width=.60\columnwidth]{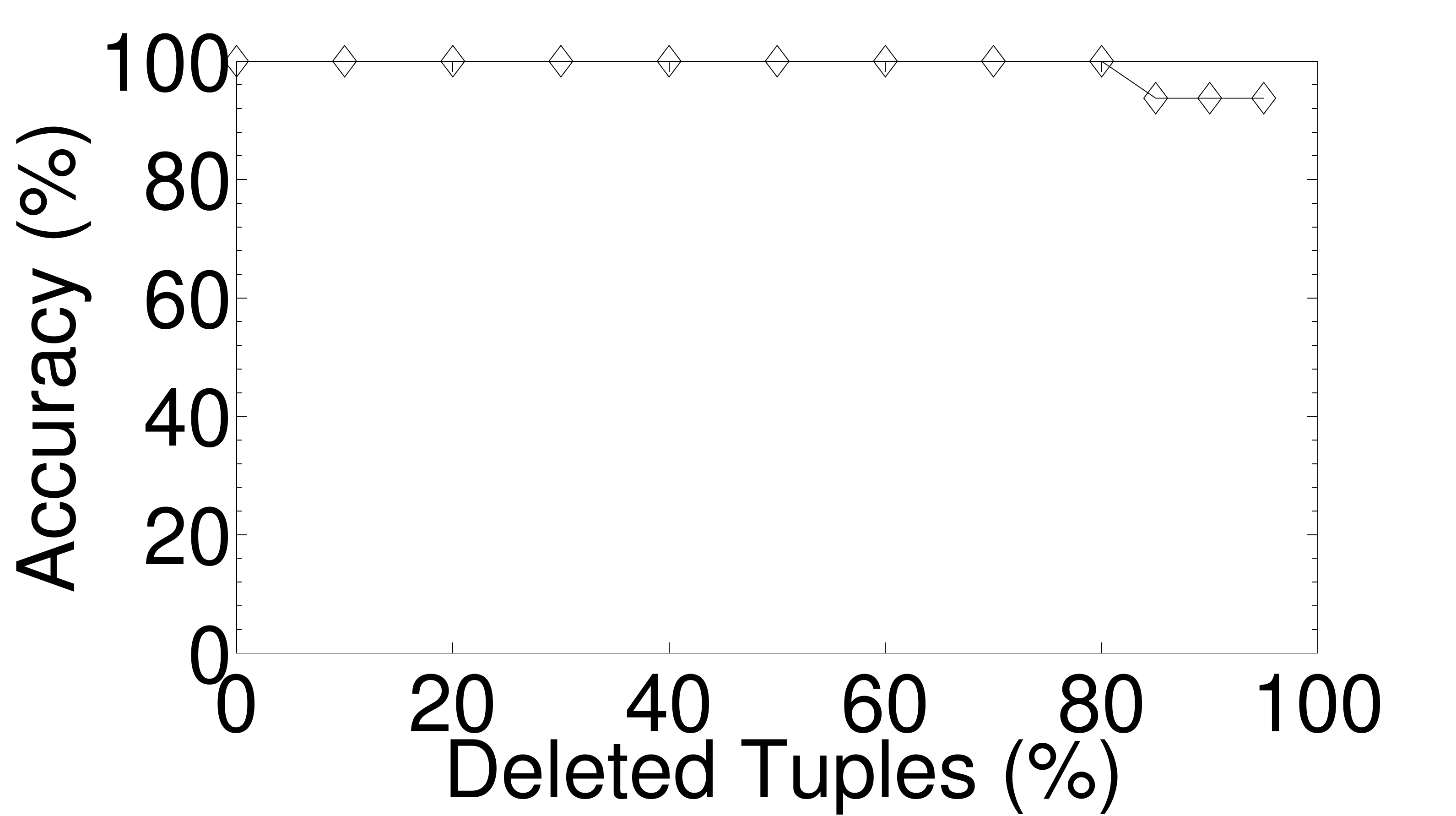} \caption{Resilience to
deletion attacks.}\label{fig:DeletionAttacks}
\end{figure}

In this attack, Mallory deletes selected  $\alpha$ tuples from the
watermarked dataset $EMR_{W}$. It is the task of the decoding
algorithm to recover the watermark from the remaining tuples in each
partition. Mallory has to preserve the usability of the dataset, so
we assume that deleting tuples does not violate the usability
constraints. In particular, if deleting $\alpha$ tuples changes the
classification potential of the feature(s) present in the EMR to an
extent where overall rank of the features is changed then the
diagnosis rules will also change; as a result, the data will not be
useful anymore for data mining. Since Mallory is unaware of other
usability constraints; therefore, to launch an attack the only
option for him is to delete $\alpha$ tuples such that the rank of
the features remain intact after the attack. As a consequence, the
data remains useful for knowledge driven CDSSs which directly depend
on the rank of the features. So, we can realistically assume that
Mallory may not be able to alter the ranks of the features while
deleting features. Therefore, our technique is able to correctly
decode the watermark as long as the features' ranks are not
disturbed. In Figure \ref{fig:DeletionAttacks}, one can see that
even when 80\% tuples are deleted, the proposed scheme is able to
recover watermark with 100\% accuracy. Similarly, when 95\% of
tuples are deleted, the scheme is still able to recover watermark
with an accuracy of 93\%. The reason is the same that it does not
use marker tuples and also does not insert watermark into specific
tuple(s).

.

%

\subsection{Alteration Attacks}
 \label{sec:AlterationAttacksResults}

\begin{figure*}[htp]\centering
\includegraphics[angle=0,
width=1\linewidth]{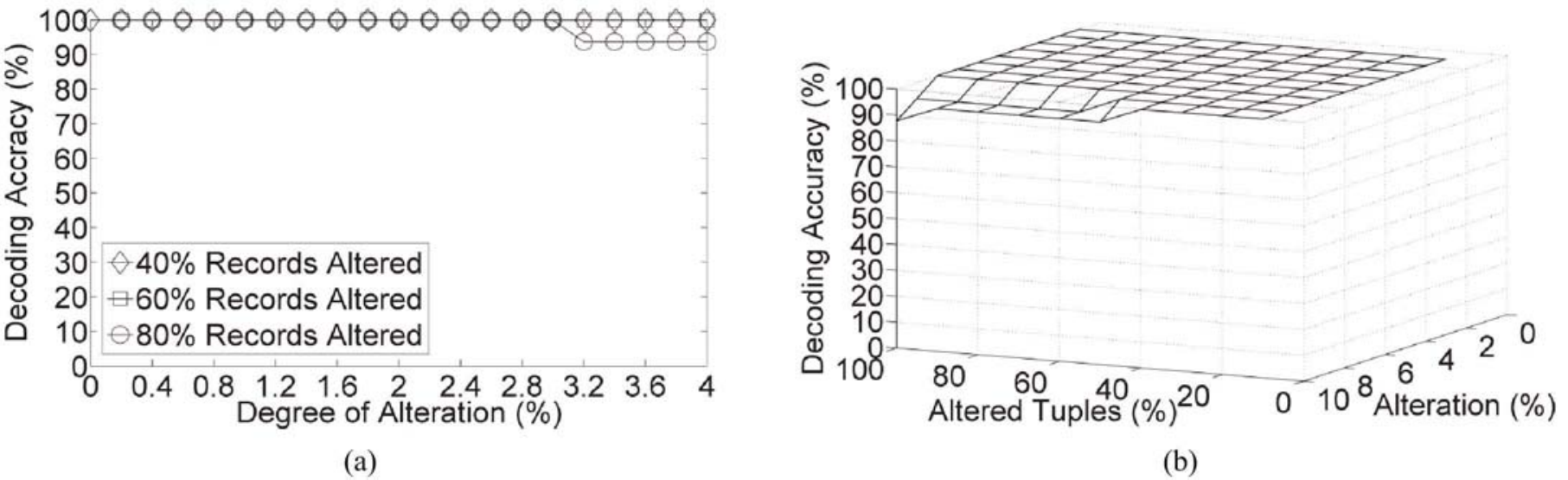} \caption{Resilience to
alteration attacks. (a) Resilience to random
($\alpha,\varrho$)-alteration attacks. (b) Resilience to fixed
($\alpha,\varrho$)-alteration attacks.
}\label{fig:ResilienceToAlterationAttacks}
\end{figure*}

Mallory can alter the value of an attribute by changing the value of
an attribute in a random or fixed manner. In case of random
alterations, Mallory selects $\alpha$ tuples and changes the values
of features in the range $[0, \pm\varrho]$; hence they are termed as
($\alpha,\varrho$ ) alteration attacks. One can see in Figure
\ref{fig:ResilienceToAlterationAttacks} that the proposed technique
successfully retrieves the watermark even when 60\% records are
altered. The watermark decoder caters for such data alterations and
hence provides robustness against such alterations.

In the second case, Mallory selects $\alpha$ tuples and alters the
value of a feature from \emph{x} to ($ x\pm\varrho $). But we see in
Figure \ref{fig:ResilienceToAlterationAttacks} that the proposed
technique is resilient to such fixed alteration attacks as well.
This is because the decoding process uses watermark decoder that
uses information-preserving watermarking parameter $\beta$.

A sophisticated attacker may try to simultaneously launch deletion,
insertion, and alteration attacks. In this case, the probability of
successful attack is $(o.5)^{\frac{(\alpha + \omega)}{2}}$, where
$\alpha$ is the number of records altered, and $\omega$ is the
number of unaltered records selected from the watermarked dataset.
We have also conducted a number of experiments by changing the
percentage of records added, deleted and altered. Our findings are
that if an attacker deletes up to 40\% (of $R$) of
 records, adds up to 50\%(of $R$) new records, and alters up to 40\%(of $R$) records,
 we are still able to recover the watermark with 100\% accuracy.

Moreover, we also study the imperceptibility of the proposed
watermarking technique. The objective of this study is to analyze
the correlation between the inserted watermark -- using our
information-preserving watermarking scheme -- and the recovered
(decoded) watermark using different decoding schemes. The study
proves that a watermark inserted with our technique could be
successfully decoded using our decoding scheme only. For this
purpose, correlation coefficient \cite{anderson1958introduction} is
used. The results tabulated in Table
\ref{table:CorrelationAmongDifferentWatermarks} are the average
values of correlation coefficient obtained from $100$ different runs
of each algorithm. It is clear that the proposed technique achieves
the best correlation coefficient through different runs of the
algorithms. Any other watermark decoding techniqe never achieves
more than $0.16$ correlation coefficient.

\begin{table}[t]
\scriptsize
  \centering
  \caption{Correlation among watermarks.}
  \label{table:CorrelationAmongDifferentWatermarks}
\begin{tabular}{|l|l|l|l|}

\hline
\textbf{l} & \textbf{Proposed Scheme} & \textbf{Shehab's Scheme}  & \textbf{Probabilistic Decoding} \\
\hline
8 bits & 1 & 0.16 & -0.07  \\
\hline
16 bits & 1 & 0.13 & -0.22 \\
\hline
32 bits & 1 & 0.12 & -0.06 \\
\hline
64 bits & 1 & -0.23 & 0.1 \\
\hline

\hline
\end{tabular}
\end{table}

\section{Conclusion}
 \label{sec:Conclusion}

In this paper, a novel information-preserving technique for
watermarking EMR is presented. The benefits of this technique are:
(1) preserving the classification potential of high ranking
features, (2) empirically selecting the length of watermark to
ensure realtime computability constraints, (3) ensuring the
usability constraints, and (4) decoding the watermark using majority
voting based on a novel watermark decoder. The benefit of using
watermark decoder is that it makes the technique resilient to
malicious attacks, even in presence of violations of few usability
constraints. We have compared the proposed technique with a recently
proposed state-of-the-art technique to show that the technique
preserves the diagnostic rules of medical data; as a result, the
misdiagnosis rate is significantly reduced. To the best of our
knowledge, no watermarking technique utilizes the information of
classification potential. Moreover, the technique is also resilient
to insertion, deletion and alteration attacks. In future, we would
like to extend the technique on non-numeric strings data and medical
images.

 \ifCLASSOPTIONcompsoc
  \section*{Acknowledgments}
\else
  \section*{Acknowledgment}
\fi

The first author is funded by Higher Education Commission (HEC) of
Pakistan through a Ph.D. fellowship under its indigenous scheme. The
data for high risk patients is collected by team members of Remote
Patient Monitoring System project
(http://rpms.nexginrc.org/index.aspx). The authors are thankful to
the management of nexGIN RC and CureMD for sharing Weka compliant
datasets -- extracted from their EMR systems -- for this research.

\ifCLASSOPTIONcaptionsoff
  \newpage
\fi

\vspace*{-2\baselineskip} \vfill
\begin{IEEEbiography}[{\includegraphics[width=1in,height=1.25in,clip,keepaspectratio]{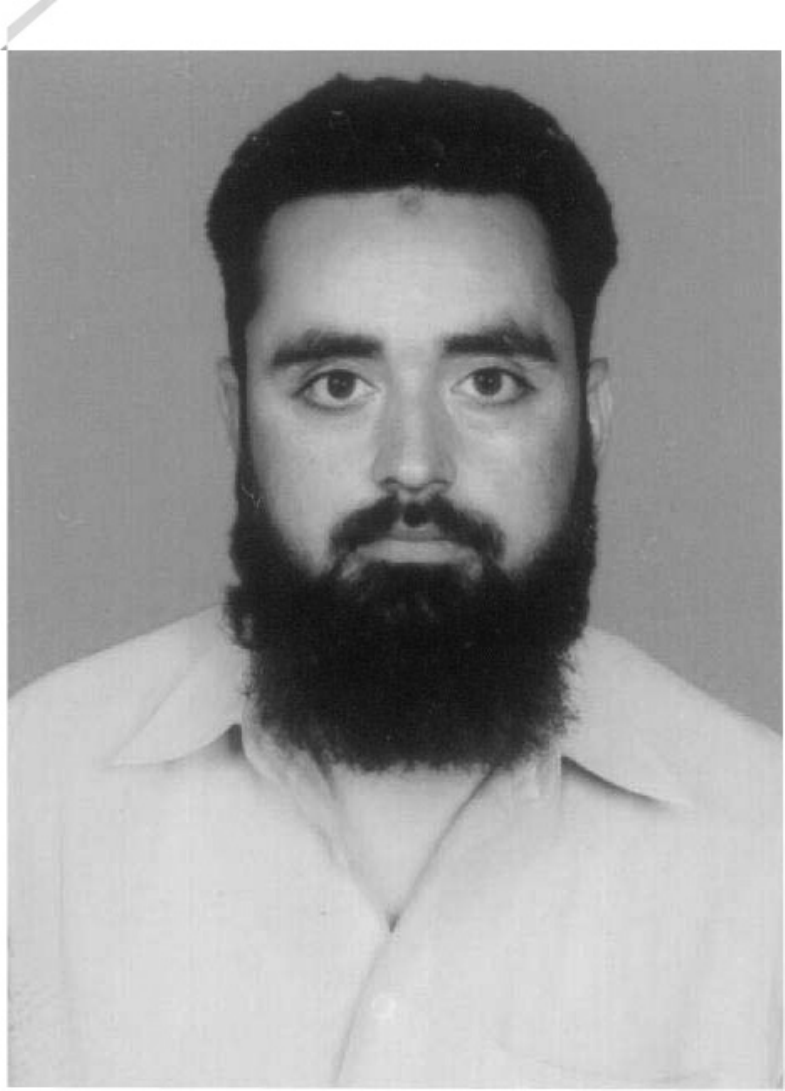}}]{M. Kamran}
 got his BS and MS degree in Computer Science in 2005 and
2008 respectively. Currently he is working as Assistant Professor of Computer Science at COMSATS Institute of Information Technology Wah Campus, Wah Cantt,
Pakistan. His research interests include the use of machine
learning, evolutionary computations, big data analytic, data security, and decision support systems.
\end{IEEEbiography}
\vspace*{-2\baselineskip}
\begin{IEEEbiography}[{\includegraphics[width=1in,height=1.25in,clip,keepaspectratio]{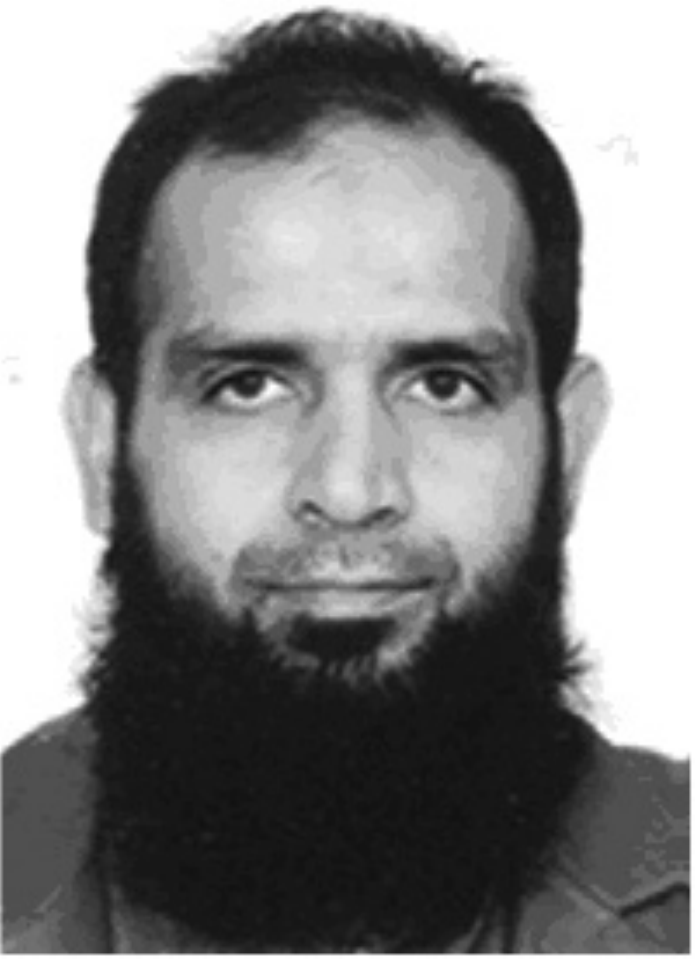}}]{Muddassar Farooq}
received his B.E. degree in Avionics Engineering from National University of Sciences and Technology (NUST), Pakistan, in 1996. He completed his M.S. in Computer Science and Engineering from University of New South Wales (UNSW), Australia, in 1999. He completed his D.Sc. in Informatics from Technical University of Dortmund, Germany, in 2006. In 2007, he joined the National University of Computer and Emerging Sciences (NUCES), Islamabad, Pakistan, as an associate professor. Currently, he is working as the Director of Next Generation Intelligent  Networks Research Center (nexGIN RC). He is the author of the book "Bee-inspired Protocol Engineering: from Nature to Networks" published by Springer in 2009. He has also coauthored two book chapters in different books on swarm intelligence. He is on the editorial board of Springer's Journal of Swarm Intelligence. He is also the workshop chair of European Workshop on Nature-inspired Techniques for Telecommunication and Networked Systems (EvoCOMNET) held with EuroGP. He also serves on the PC of well known EC conferences like GECCO, CEC, ANTS. He is the guest editor of a special issue of Journal of System Architecture (JSA) on Nature-inspired algorithms and applications. His research interests include agent based routing protocols for fixed and mobile ad hoc networks (MANETs), nature inspired applied systems, natural computing and engineering and nature inspired computer and network security systems, i.e. artificial immune systems.

\end{IEEEbiography}

\end{document}